\documentclass[12pt,preprint]{aastex}

\slugcomment{Accepted by Astrophys. J}

\shorttitle{Explosive Nucleosynthesis Associated with GRBs}
\shortauthors{Nagataki et al.}

\begin{document}

\title{Explosive Nucleosynthesis Associated with Formation of Jet-induced
GRBs in Massive Stars}


\author{S. Nagataki\altaffilmark{1}, A. Mizuta\altaffilmark{2}, 
S. Yamada\altaffilmark{3}, H. Takabe\altaffilmark{2}, and K. Sato\altaffilmark{1,4}}
\email{nagataki@utap.phys.s.u-tokyo.ac.jp}

\altaffiltext{1}{
Department of Physics, School of Science, University
of Tokyo, Tokyo 113-0033, Japan
}
\altaffiltext{2}{
Institute of Laser Engineering (ILE), Osaka University, Yamadaoka, Suita,
Osaka 565-0871, Japan
}
\altaffiltext{3}{
Science and Enjineering, Waseda University, Okubo 3-4-1, Shinjyuku,
Tokyo 169-8555, Japan
}
\altaffiltext{4}{
Research Center for the Early Universe, University of Tokyo,
Tokyo 113-0033, Japan
}


\begin{abstract}
We perform 2-dimensional relativistic hydrodynamical simulations
in the context of collapsar model. Calculations of explosive nucleosynthesis
are also accomplished. We investigate the influence
of the structure of the progenitor and energy deposition rate on the resulting
explosive nucleosynthesis, assuming that $\rm ^{56}Ni$ is mainly synthesized
in the jet launched by the neutrino heating.
We show the amount of $\rm ^{56}Ni$
is very sensitive to the energy deposition rate.
Thus we conclude that it is quite natural not to detect an underlying supernova
in some X-ray afterglows as in GRB 010921.
We also point out the possibility that the relative
abundance of the elements with intermediate mass number
such as Si and S in the X-ray afterglow
of GRB 011211 may be naturally explained if the energy
deposition rate at the central engine is relatively long because little
amount of $\rm ^{56}Ni$ should be synthesized under such an environment. 
If this discussion is true, there should be correlation 
between the line features in the X-ray afterglow and duration of the GRB.
It should be noted that the duration of GRB 011211 is
270 seconds, making it the longest burst ever observed by Beppo-SAX although
it suffers from the effect of red-shift ($z_{\rm host}=2.14$), which
supports our conclusion. Our results also suggest that
the type I collapsar model in which the energy deposition rate is relatively
low ($\dot{E} \sim 10^{51}$ erg s$^{-1}$) might have difficulty in reproducing
the observed amount of $\rm ^{56}Ni$ in a hypernova such as SN 1998bw.
This means that the mechanism of the central engine of a hypernova 
accompanying GRB may be constrained by the discussion of explosive
nucleosynthesis. 
\end{abstract}




\section{Introduction}\label{intro}

A revolution in our understanding of gamma-ray bursts (hereafter GRBs)
occurred in 1997, following
accurate localization of GRBs by BeppoSAX~\cite{boella97} and the discovery
of the afterglow phenomenon. Some GRBs and afterglows are known to have
extragalactic origins, which are nicely accounted for by fireball
models~\cite{meszaros97}, although the
origin of the fireball is not yet clear.

However, there has been growing evidence linking GRBs to massive
stars. The host galaxies of GRBs are star-forming galaxies
and the position of GRBs appear to trace the blue light of young
stars~\cite{tsvetkov01, bloom02, floch03}. Some of the host galaxies appear to
be dusty with star-formation rates comparable to ultra-luminous infrared
galaxies~\citep{berger01,frail02}. On smaller scales, there is mounting
evidence of association with regions of high ambient
density
\citep{galama01,harrison01}
and the so-called dark GRBs arise in or behind regions of high
extinction~\cite{piro02}.

Moreover, there has been tantalizing evidence for the existence of
an underlying supernova
(SN). The first association of a cosmologically distant GRB with the death of
a massive star was found for GRB 980326, where a clear excess of emission
was observed over and above the rapidly decaying afterglow component.
This late-time bump was interpreted as a contribution of an underlying
SN~\cite{bloom99} since, unlike the afterglow, the bump was very red.
GRB 970228, also with an intermediate-time bump and characteristic SN
spectral rollover, is another good candidate~\citep{reichart99,galama00}.

It was also reported that there seems to be a physical connection
between GRB 980425 and SN 1998bw~\cite{galama98}. They discovered an
optical transient within the error box of BeppoSAX Wide Field Camera for
GRB 980425. It should be noted that this SN is categorized as a new
type of SN (i.e. hypernova) with large kinetic energy ($\sim 10^{52}$ ergs),
nickel mass ($\sim 0.5M_{\odot}$), and luminosity~\citep{iwamoto98,woosley99}.
In the analysis of an underlying supernova mentioned above, the light curve
of SN1998bw is usually used for the SN template.

Generally speaking, it is considered that it is
too difficult to realize a GRB from a
death of a massive star, because a fire-ball is required to be composed of
enormous radiation energy ($\sim 10^{51}$ erg) with very small baryon number
($\sim 10^{-6}M_{\odot}/m_p$, where $m_p$ is the proton 
mass)~\cite{meszaros97}. 
One of the most famous models to realize a GRB and large explosion energy
from a death of a massive star is the collapsar
model~\cite{woosley93,macfadyen99}.
Collapsar is defined as a massive star whose iron core has collapsed to a
black hole that is continuing to accrete envelope matter
at a very high rate. 
Woosley also pointed out that there will be two types
for collapsars. One (type I collapsar) is that the central core
immediately forms a black hole with an accretion disk. In the other case
(type II collapsar), the central core forms a neutron star at
first, but the neutron star collapses to a black hole with an
accretion disk due to the continuous fall back. 
In both types, it is pointed out that an accretion disk should
be formed around the central compact object and neutrinos should be
emitted from the innermost region of the accretion disk through the
electron (positron) capture and electron-positron pair
annihilation~\cite{berezinskii87, macfadyen99}. Finally, a strong jet with large radiation
energy and small baryon number, which is required to produce a GRB, is
generated around the polar region due to the pair-annihilation of neutrinos
that come from the accretion disk and/or MHD processes.
In fact, it is reported that the narrow line of oxygen
and broad lines of iron in SN 1998bw can be well reproduced
when jet-like explosion is assumed~\cite{maeda02}.

One may have thought that the system of a GRB has been well known
and there is no question about that. That is, a GRB is born from
a death of a massive star which collapses and makes a black hole
at the center forming a strong jet with a large photon to baryon ratio,
large explosion energy of order $\sim 10^{52}$ ergs, and
a great amount of $\rm ^{56}Ni$ ($\sim 0.5M_{\odot}$) that makes
the underlying SN so luminous that it is categorized as a hypernova.
However, there are some important problems with GRBs.

First, it is reported that no coincident SN is detected
at 99.7$\%$ confidence by Hubble
Space Telescope in GRB 010921 to a limit 1.34 mag fainter than SN 1998bw.
The host galaxy's redshift is relatively
low ($z$=0.451), which made intensive multi-color observations possible.
It is also pointed out that the observed luminosity of SN 1998bw may be
reproduced with a smaller amount of $\rm ^{56}Ni$ than that required in the
spherically-symmetric models by invoking the
angle-dependent luminosity expected from asymmetric explosion~\cite{wang02}.
Thus it seems to be still uncertain whether a massive star,
that makes a GRB,
always produces much of $\rm ^{56}Ni$.

Second, there are some reports on the strong iron K$\alpha$ emission
lines~\citep{piro98,piro00,antonelli00,yoshida01}
in X-ray afterglows. Moreover, there is a report
of emission lines of Mg, Si, S, Ar, and Ca with an outflow velocity of
order 0.1$c$ in the X-ray afterglow of GRB 011211~\cite{reeves02},
although it seems to be still controversial~\cite{rutledge02}.
At present, the origin of these lines is still open to arguments.
We should investigate whether these features can be explained in the
context of collapsar model.

Third, it is an open question whether $\rm ^{56}Ni$ is synthesized
enough to explain the light curve of SN 1998bw in the collapsar model.
As explained below, there is an inconsistent treatment between
the collapsar model and the hypernova model.
In a simulation of the type I collapsar, the mass accretion
rate is estimated to be $\sim 0.07M_{\odot}$ s$^{-1}$ in the phase of jet
formation~\cite{macfadyen99}, and the timescale of neutrino emission from
the accretion disk is also estimated to be about 10 seconds since
the temperature of the inner-most region of the accretion disk
becomes lower than 1 MeV after that
and electron-positron pair creation becomes
impossible. As a result, the neutrino emission from the accretion disk
ceases~\citep{macfadyen99,nagataki02}. In fact, the thermal energy
is usually injected around the polar region at a rate $\dot{E} \sim 10^{51}$
erg s$^{-1}$ for about 10 seconds in numerical simulations, which makes
the total explosion energy $\sim 10^{52}$ ergs~\citep{macfadyen99,aloy00}. 
On the other hand, in a simulation of explosive nucleosynthesis in the context
of a hypernova, total explosion energy
is put at once below the mass cut that divides the ejecta from the
collapsing core as an initial condition~\cite{maeda02}.
In their simulations, considerable fraction ($\ge 0.5$) of the explosion
energy is given as kinetic energy, as opposed to the simulation of collapsars.
Also, the progenitor of a hypernova is assumed to be spherical.
Thus it should be necessary whether these simulations are consistent
with each other or not. That is, we should investigate whether a
collapsar model can explain the phenomenon of SN 1998bw as in the hypernova
model.

It should be noted here that the accretion disk around the central black hole
may be the site where $\rm ^{56}Ni$ is
synthesized~\citep{woosley99,macfadyen99,woosley02,pruet03}.
Pruet et al. (2003) calculated electron fractions in the
accretion disk and found the proper condition to produce $\rm ^{56}Ni$
in the disk. They also pointed out the possibility that a considerable
amount of $\rm ^{56}Ni$ can be driven by the disk wind. Although their scenario
is widely supported to be very promising, the quantative evaluation of the
yields has yet to be done.
In this study,
we consider another possibility that $\rm ^{56}Ni$ is synthesized
in the jet, as investigated in Maeda et al. (2002). Thus reader should
be reminded that this is our assumption in this study. We also have to note
that the opening angle of the jet launched by the neutrino heating has
to be large in our scenario, as shown in the following sections, to produce
a large enough amount of $\rm ^{56}Ni$ to explain the luminosity of the
hypernova such as SN 1998bw. Such a wide jet will not make a GRB because a
bulk Lorentz factor can not be so large as to realize a fireball.
Thus, in our scenario, another narrow jet produced possibly
by MHD effect ir as a small part of the wider jet will be needed to
realize a GRB. In this study, however, we do not consider it explicitly,
since, if any, it should have little influence on the product
of explosive nucleosynthesis because the baryon mass in the narrow jet should
quite small.

In this study, explosive nucleosynthesis is calculated in the context of
type I collapsar model. At first, a progenitor collapses and
forms a black hole at the center. Next, thermal energy is injected
around the polar region to make a jet. Finally, explosive nucleosynthesis
is calculated by the post-processing method. We investigate the influence
of the progenitor structure and energy deposition rate on the results of
explosive nucleosynthesis. We show the amount of $\rm ^{56}Ni$
is very sensitive to the energy deposition rate. A large amount
of $\rm ^{56}Ni$
is produced when the energy deposition rate is high and vice versa.
Thus we conclude that it is quite natural not to detect an underlying SN
in some cases. In such cases, we consider that the energy deposition
rate was relatively low and a small amount of $\rm ^{56}Ni$ was
synthesized. We also give a discussion on the 
chemical composition of the ejecta of a collapsar and observed line
features in some X-ray afterglows. We point out the possibility
that the variation of chemical composition
in the X-ray afterglow may be naturally explained by the variation
of the energy deposition rate at the central engine.
By this discussion, we can predict that
there should be correlation between the line features in the X-ray
afterglow and duration of the GRB.
We also compare the calculated amount of $\rm ^{56}Ni$ with
the observed value in SN 1998bw and investigate which model is best.
We show that the type I collapsar
model in which the energy deposition rate is relatively low
($\dot{E} \sim 10^{51}$ erg s$^{-1}$) may have difficulty in reproducing the
observed amount of $\rm ^{56}Ni$ of SN 1998bw and GRB 980425 and
some improvements should be required to explain these observations 
in the context of the type I collapsar model.
This means that the mechanism of the
central engine of a hypernova and a GRB is
constrained by the discussion of explosive nucleosynthesis.

In section~\ref{method}, the method of calculation is explained in detail.
Results are shown in section~\ref{results}. Discussions
are presented in section~\ref{discussion}. Conclusion is given
in section~\ref{conclusion}.

\section{Method of Calculation}\label{method}

In this section, we present the method of the simulations in this
study. It is true that there are some assumptions and simplifications
in our simulations. In particular, the effects of gravitation and rotation
are not included in this study. In order to realize the jet-induced explosion,
we prepare an asymmetric progenitor as an initial condition following the 
radial infall with the infall timescale depending on the zenith
angle. In this way, we can mimic an asymmetric progenitor models
of MacFadyen and Woosley (1999). After that, we realize
the jet-induced explosion by injecting the thermal energy around the
polar region in the same way as MacFadyen and Woosley (1999) did.
Of course, this needs improvement in the future.
We explain the assumption and simplification in more detail
in the following subsections.

\subsection{Hydrodynamics} \label{hydro}
\subsubsection{The Scheme} \label{scheme}

We performed 2-dimensional hydrodynamic calculations using a
spherical coordinate taking account of special relativity.
The calculated
region corresponds to a quarter of the meridian plane under the
assumption of axial-symmetry and equatorial symmetry. The number of
mesh points is 500 in the radial direction and 50 in the
angular one for the calculation of explosive nucleosynthesis.
The radial mesh size is set to be constant in space, 2$\times 10^7$ cm.
With this resolution, the structure of the central accretion disk shown
by MacFadyen and Woosley (1999) can be marginally resolved. 
We did not perform 3-dimensional hydrodynamic
calculations simply for saving CPU time and memory. 
We employed an approximate Riemann solver, Maquina's flux
formula~\cite{donat96,donat98}, for special relativistic hydrodynamic
equations.
The code is of first order accuracy both spatially and temporally.
In this study, we assume that the gas is ideal with the 
equation of state $p=\rho\epsilon (\gamma -1)$,
where $p$, $\rho$, $\epsilon$, and $\gamma$ are pressure, rest mass density,
specific internal energy and adiabatic exponent
(=4/3 constant in this paper), respectively.
Since the hydrodynamics code is Eulerian,
we use the test particle
method~\cite{nagataki97}
in order to obtain the informations on the time evolution of the
physical quantities along the fluid motion, which are then used
for the calculations of the explosive nucleosynthesis.
Test particles are
scattered in the progenitor and are set
at rest at first. They move with the local fluid
velocity at their own positions after the passage of the shock wave. 
The temperature and density that each test particle
experiences at each time step
are preserved. This is the test particle method we use.
See Nagataki et al. (1997) and Nagataki et al. (1998) for more details.

Calculations of hydrodynamics and explosive nucleosynthesis are
performed separately, since the entropy produced during the explosive
nucleosynthesis is much smaller ($\sim$ a few$\%$) than that generated by
the shock wave.
In calculating the total yields of elements, we
assume that each test particle has its own mass determined
from their initial distribution so that their sum
becomes the mass of the layers where these are scattered.
It is also assumed
that the nucleosynthesis occurs uniformly in each mass element.
These assumptions will be justified since the movement of the test
particles is not chaotic (i.e. the distribution of test particles at
the final
time still reflects the given initial condition) and
the intervals of test particles are sufficiently narrow to give a smooth
distribution of the chemical composition in the ejecta.
The number of the test particles are 42000. The test particles
are distributed uniformly within the radius of 7$\times 10^9$ cm.

\subsubsection{Initial and Boundary Conditions} \label{initial}

As for the progenitor of SN 1998bw, it is thought to
have had the mass $\sim 40 M_{\odot}$ in the main-sequence
stage~\citep{iwamoto98,woosley99} and had $\sim$
16$M_{\odot}$ helium core~\cite{iwamoto98}.
In this study, the presupernova model obtained
from the evolution of 16 $M_{\odot}$ helium core~\cite{nomoto88}
is used for the calculation of explosive nucleosynthesis.

\begin{figure}
\plotone{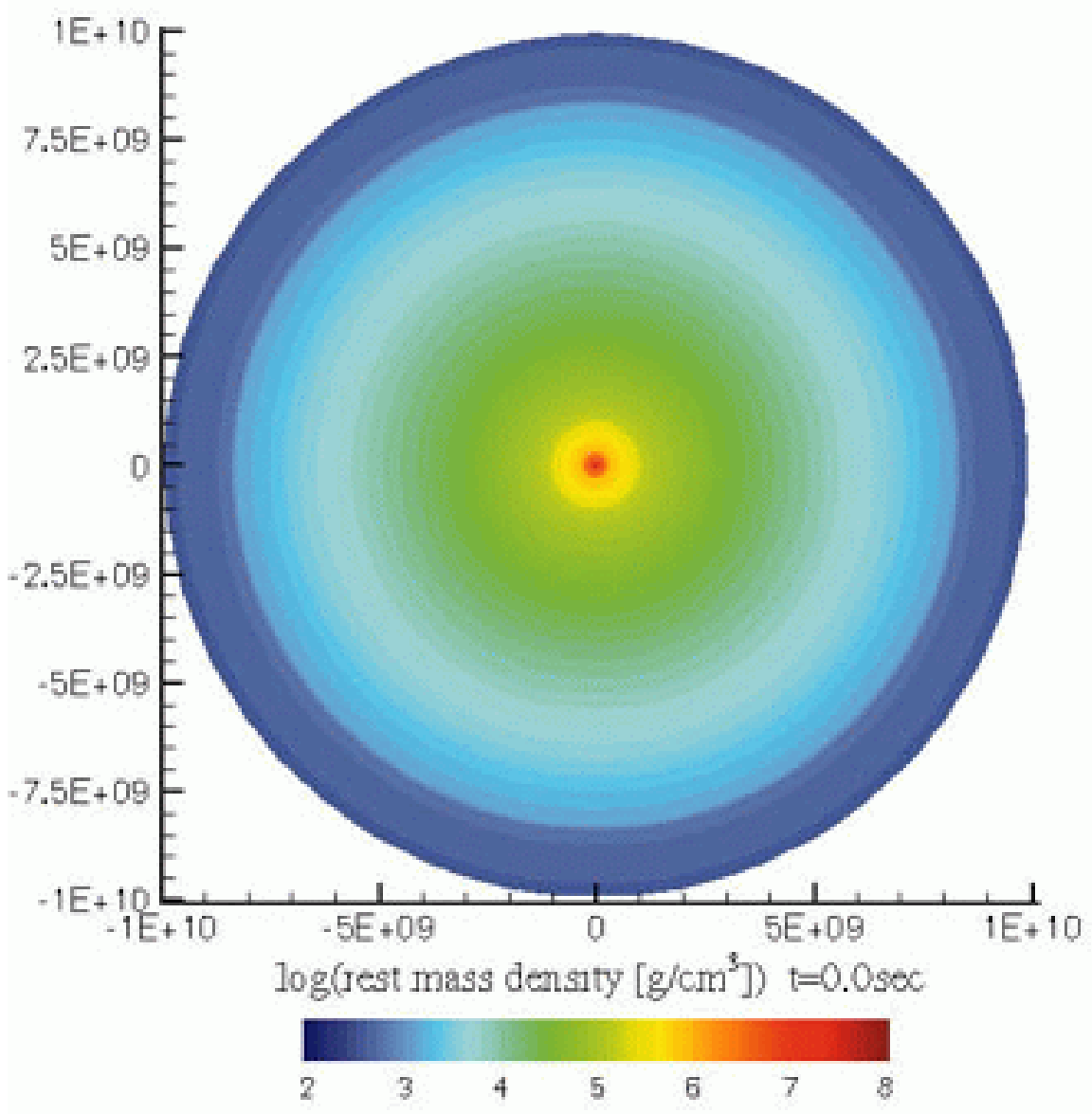}
\caption{Initial density contour for the spherical model. The free-fall
time is determined so that the mass of the central bleck hole
becomes 3$M_{\odot}$. 
\label{fig1}}
\end{figure}

\begin{figure}
\plotone{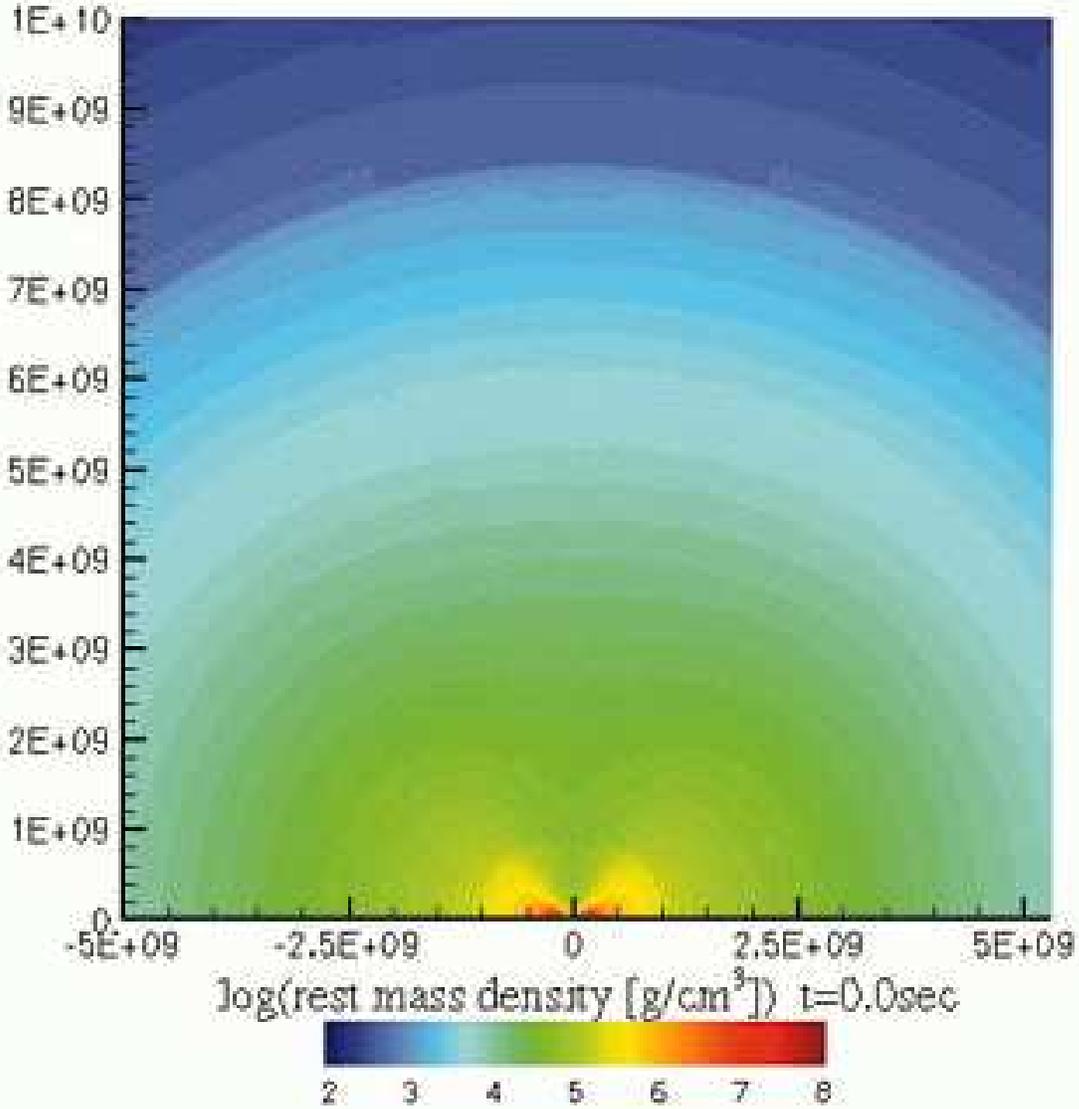}
\caption{Same as Fig.1 but for the asymmetric model and only the northern
heimsphere is shown.
The asymmetric model is made by allowing only the radial infall and
assuming that the infall timescale depends on the zenith angle.
The infall time is determined so that the central mass of the black hole
becomes 3$M_{\odot}$ and the density structure resembles to the result of
MacFadyen and Woosley (1999). 
\label{fig2}}
\end{figure}

As explained above, we make an asymmetric progenitor from the spherical
presupernova model by assuming that matter free-falls radially during
the times which depend on the zenith angle. The infall time is determined
so that the mass of the central black hole becomes 3$M_{\odot}$ 
at appropriate time and the
asymmetric progenitor resembles to the result of MacFadyen
and Woosley (1999) in which the effects of gravitation and rotation are
taken into account (see their figures 7 and/or 9). 
The adopted function for the free-fall timescale is 
\begin{eqnarray}
t_{\rm fall} = c_1 \exp(-(\theta/\sigma)^2) \;\;\; \rm [sec]
\label{eqn1}
\end{eqnarray}
where $\theta$ is the zenith angle (in units of radian), and $c_1$ and $\sigma$
are set to be 1.0 and $\pi/4.6$, respectively.
After the radial infall, we can determine the Schwarzschild radius and
the baryon mass of the black hole. As mentioned above, the parameters
in Eq.~(\ref{eqn1}) are so shosen to give the baryon mass
of the central black hole 3$M_{\odot}$.
As will be explained in section~\ref{discussion}, the neutrino heating
is effective as long as the mass of the central black hole is less
than$\sim 4M_{\odot}$.
Hence, if the mass accretion
rate is $\sim 0.1M_{\odot}$ s$^{-1}$ and the duration of the
neutrino heating is $\sim 10$ s as inferred from the result of MacFadyen
and Woosley (1999), the mass of the central black
hole will be $\sim 3M_{\odot}$ at the time of the jet launch. 
In fact, MacFadyen and Woosley (1999) launched the jet when the mass of the
central black hole becomes 3.5$M_{\odot}$ in their calculation, which
is consistent with our treatment. Aloy et al. (2000) also
launched the jet when the mass of the central black hole becomes
3.762$M_{\odot}$ in their calculation.  The density, on the other hand, at the
innermost region around the polar axis is about $10^{6}$ g cm$^{-3}$ and
that around the equatorial plane is about $10^{8}$ g cm$^{-3}$ in our model,
which is
similar to the result of MacFadyen and Woosley (1999). 
This indicates that our prescription Eq.~(\ref{eqn1}) approximately
reproduces the more realistic models of MacFadyen and Woosley (1999).
It is noted that the accretion disk obtained numerically by MacFadyen
and Woosley (1999) well matches the steady state disk (slim disk) model 
for 3$M_{\odot}$ Schwarzschild black hole (kerr parameter $a=0$),
viscosity parameter $\alpha=0.1$, and accretion rate of 0.1$M_{\odot}$
s$^{-1}$~\cite{popham99}. Since we here mimiced the numerical
result of MacFadyen and Woosley
(1999), the situation we consider corresponds to the accretion disk
around the black hole with the parameters mentioned above. 
The dependence of the explosive nucleosynthesis on different
initial conditions
will be reported in the forth-coming paper. 
For comparison, we also prepare a spherical model in which the
free-fall time does not depend on the zenith angle.
The mass of the central black hole is again set to be 3$M_{\odot}$.
The density contours of the progenitors in this
study are shown in Figs.~\ref{fig1} and~\ref{fig2}, with
the former corresponding
to the spherical model and the latter to the asymmetric model.
In the asymmetric model, the upper hemisphere is shown.

\begin{figure}
\plotone{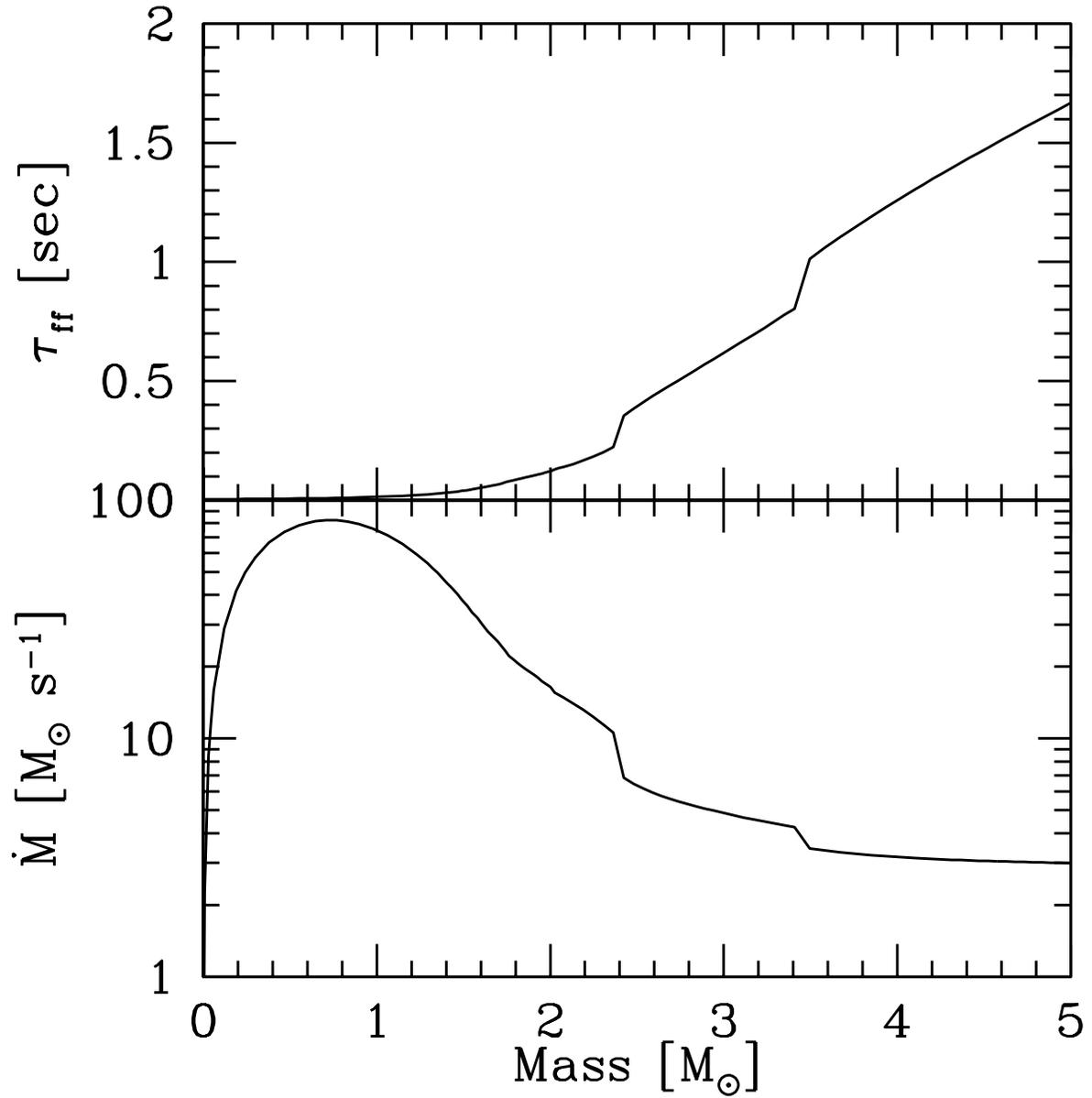}
\caption{Upper: free-fall timescale as a function of the mass coordinate.
Lower: mass accretion rate in the case of free-fall. 
\label{fig3}}
\end{figure}

It will become useful to evaluate the free-fall timescale
in the central region of the progenitor and mass accretion rate assuming
the free-fall. The free-fall timescale can be estimated as~\cite{woosley86}
\begin{eqnarray}
\tau_{\rm ff} = \frac{1}{\sqrt{24 \pi G \rho}   } \sim \frac{446}{\rho}
\;\;\; \rm [sec].
\end{eqnarray}
The mass accretion rate in the case of free-fall can be also estimated
using the mass coordinate and $\tau_{\rm ff}$ as
\begin{eqnarray}
\dot{M}(r) = \frac{M_r}{\tau_{\rm ff}} \;\;\; \rm [\it M_{\odot} \rm 
\; sec^{-1}],
\end{eqnarray}
where $M_r$ is the mass coordinate which means the enclosed mass within
the radius $r$ from the center of the progenitor. In Fig.~\ref{fig3},
we show the estimated free-fall timescale and mass accretion rate
in the case of free-fall as a function of the mass coordinate.
We can check from the figure that the mass of the central black hole
becomes $\sim 3M_{\odot}$ in 1 second in the case of free-fall.
As briefly explained in section~\ref{intro}, the neutrino emission
from the accretion disk around the black hole ceases when the temperature
of the inner-most region of the accretion disk becomes lower than 1 MeV.
This corresponds to the time when the mass of the black hole becomes $\sim
4M_{\odot}$ (see section~\ref{discussion} and Fig.~\ref{fig12}).
Thus we can find from the figure that the energy deposition 
from the accretion disk ceases within 2 seconds in the case of free-fall. 
Further, we can estimate the thermal energy deposition rate
due to the pair annihilation of (anti-)neutrinos as a function of the
mass accretion rate, which we will discuss in section~\ref{discussion}.

After making the asymmetric progenitor,
the thermal energy is injected at the innermost grid
around the polar axis at a rate $\dot{E}=10^{51}$ ergs s$^{-1}$, which
is a common treatment for the calculations of
collapsars~\citep{macfadyen99,aloy00}.
We do not consider explicitly how infalling matter are inverted.
Instead, we assume that the momentum balance is achieved when the
mass of the black hole becomes 3$M_{\odot}$.
After the launch of the jet, the thermal pressure dominates the ram
pressure since the accretion energy has declined to about a few
times 10$^{50}$ erg s$^{-1}$ at that time~\cite{macfadyen99}.
For comparison, we also perform
the computations in which the total explosion energy (= $10^{52}$ ergs)
is put instantaneously at the innermost grid around the polar axis.
The models investigated in this study are summarized in
Table~\ref{tab1}. In Models Sa/Aa, the spherical/asymmetric structure is
adopted as a progenitor into which thermal energy is injected at a rate
$\dot{E}=10^{51}$ ergs s$^{-1}$. In Models Sb/Ab, the total explosion
energy is deposited instantaneously to the spherical/asymmetric
structure.

As for the boundary condition, the reflective boundary is adopted
on the equatorial plane and symmetry axis for simplicity. It is true that
some fraction of explosion energy and matter falls into the black hole.
However, we do not observe these energy and matter and are not interested
in their chemical composition, either. Hence we consider that the deposited
energy in this study corresponds to the observed explosion energy, $10^{52}$
erg. 

\subsection{Nuclear Reaction Network} \label{nuclear}

Since the chemical composition behind the shock wave is not in nuclear 
statistical equilibrium, the explosive nucleosynthesis has to be
calculated using the time evolution of $(\rho,T)$ and a nuclear
reaction network. We obtain $(\rho,T)$ comoving with the matter
by means of the test particle
method mentioned in subsection~\ref{scheme}. The nuclear reaction network
contains 250 species (see Table~\ref{tabnucl}). We add some species around
$\rm ^{44}Ti$ to Hashimoto's network that contains 242
nuclei~\cite{hashimoto89}, although it turned out that the result was not
changed essentially by the addition.

\placetable{tabnucl}

\section{Results}\label{results}

\begin{figure}
\plottwo{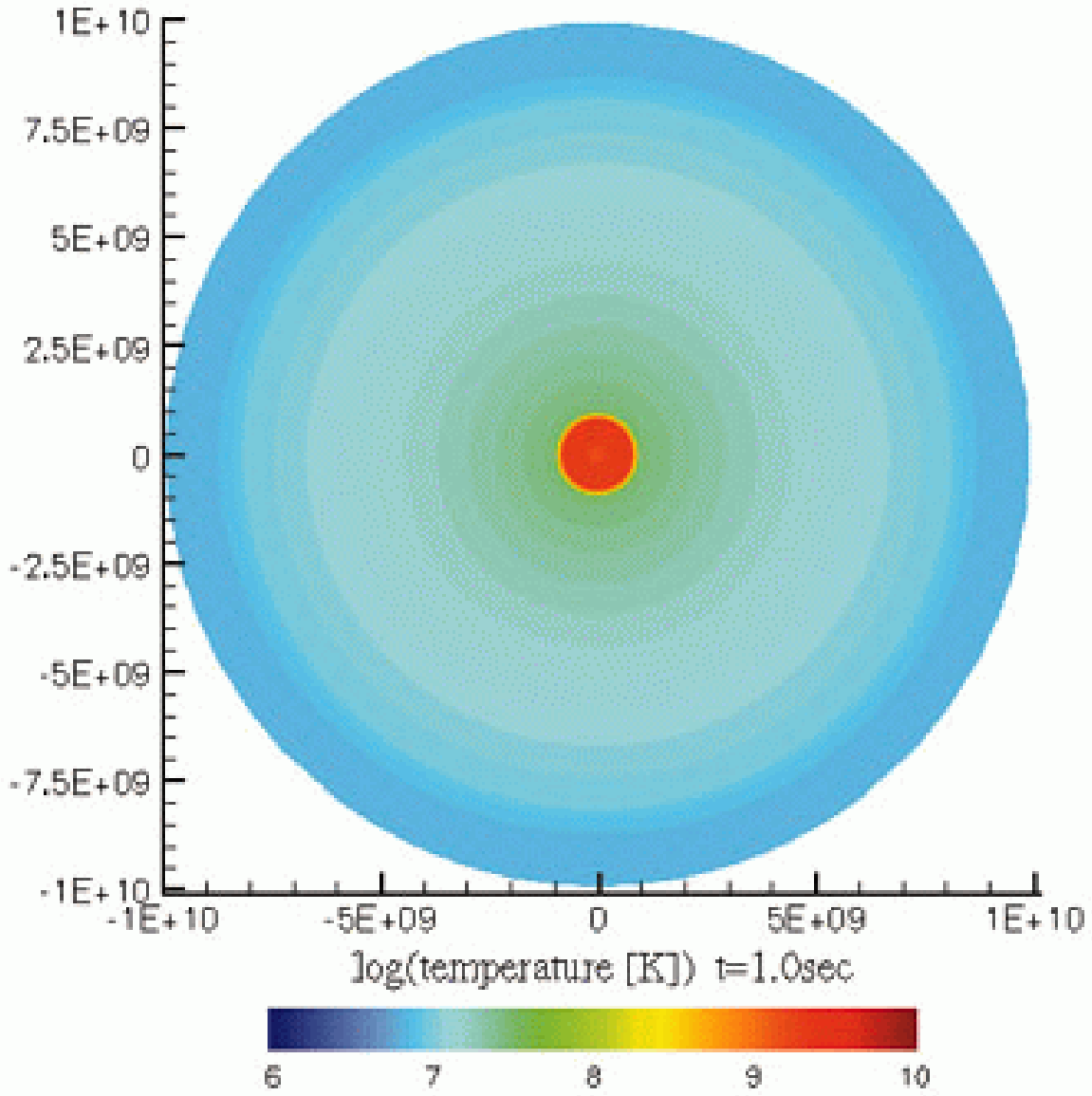}{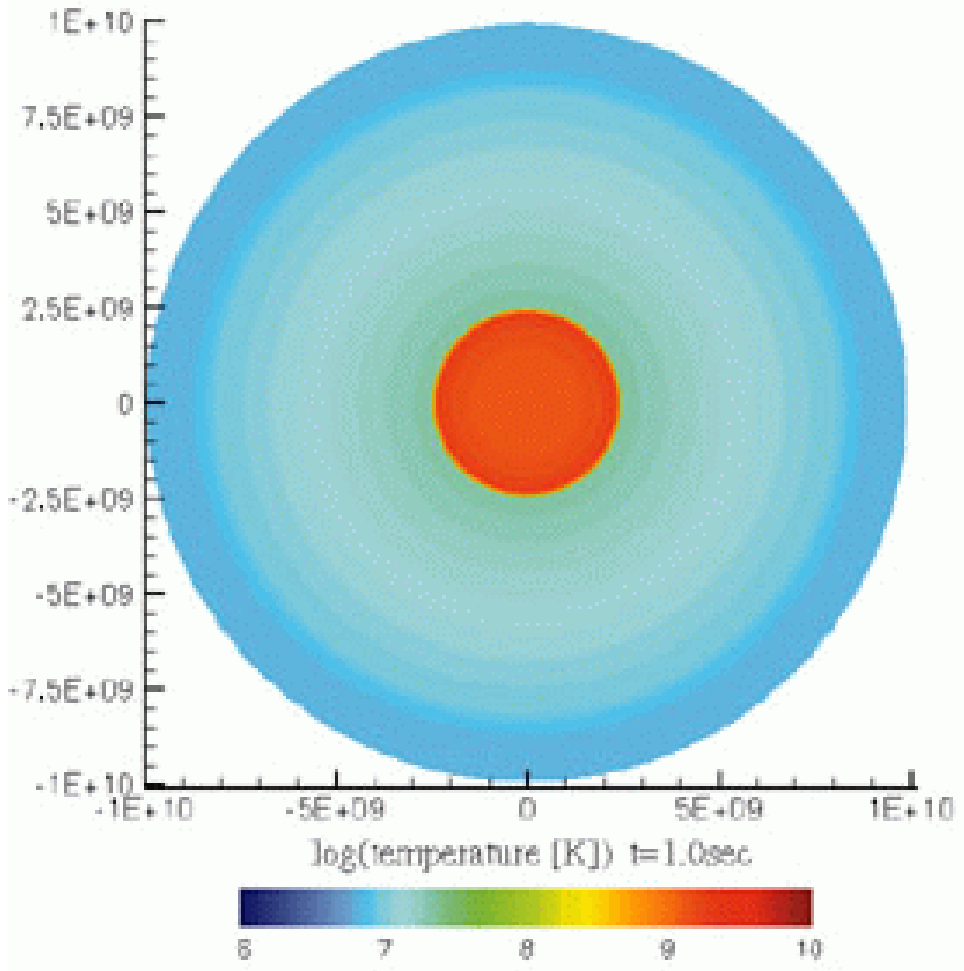}
\caption{Contours of temperature at $t=1.0$ sec. Left pannel corresponds
to Model Sa and right pannel corresponds to Model Sb.
\label{fig4}}
\end{figure}

\begin{figure}
\plottwo{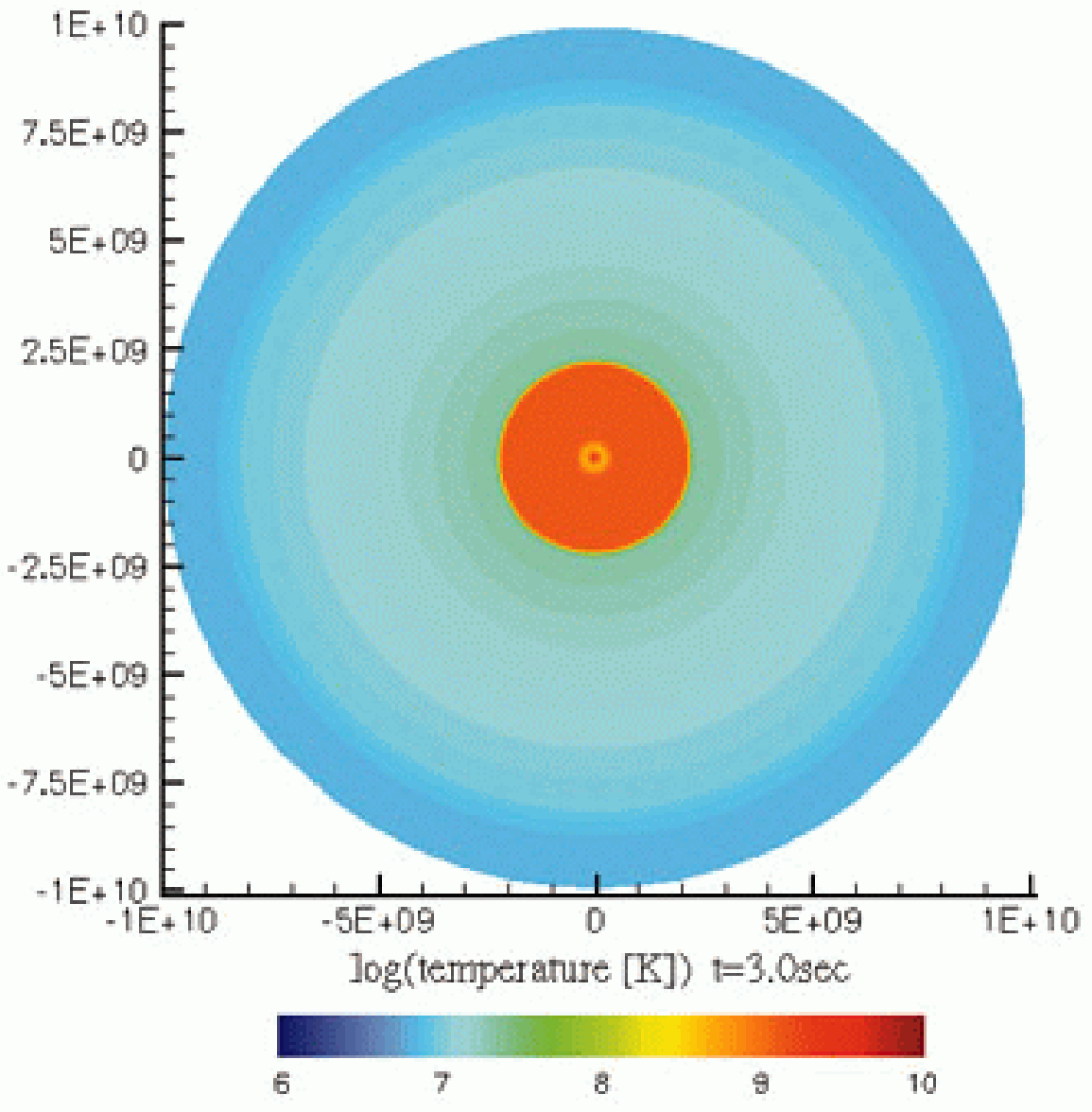}{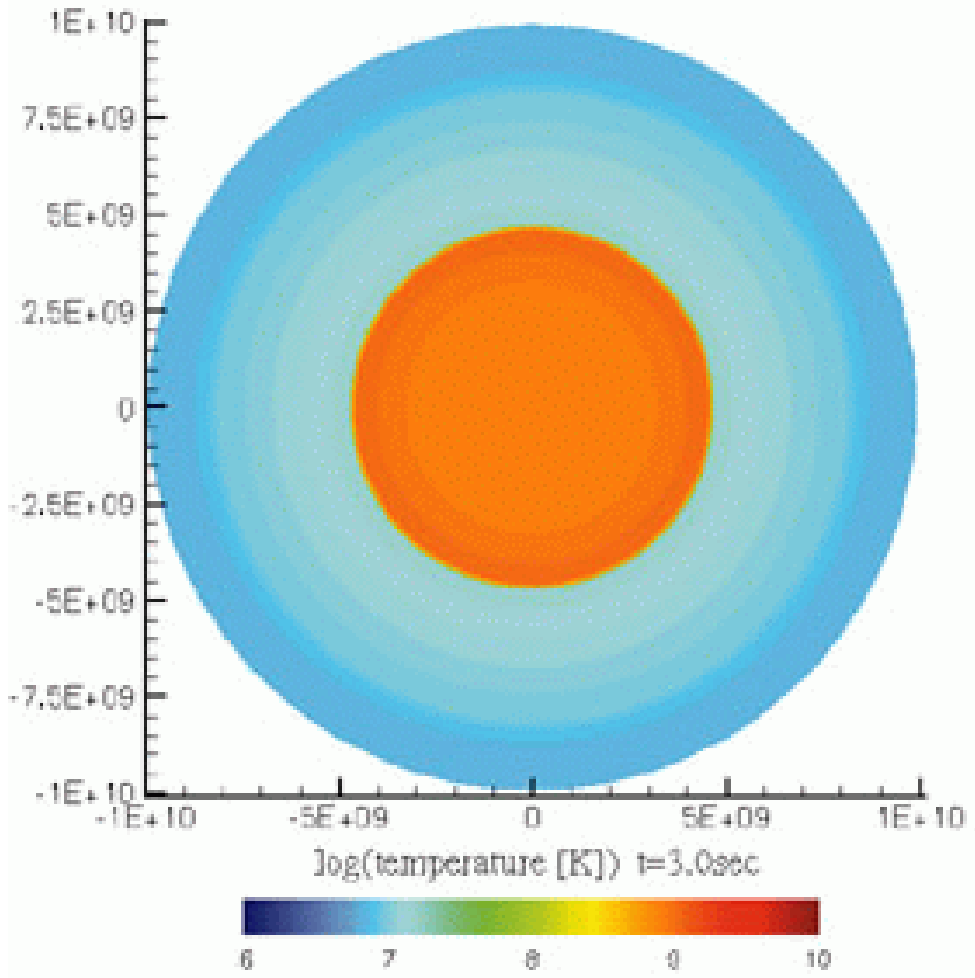}
\caption{Same as Fig.4, but for $t=3.0$ sec.
\label{fig5}}
\end{figure}

\begin{figure}
\plottwo{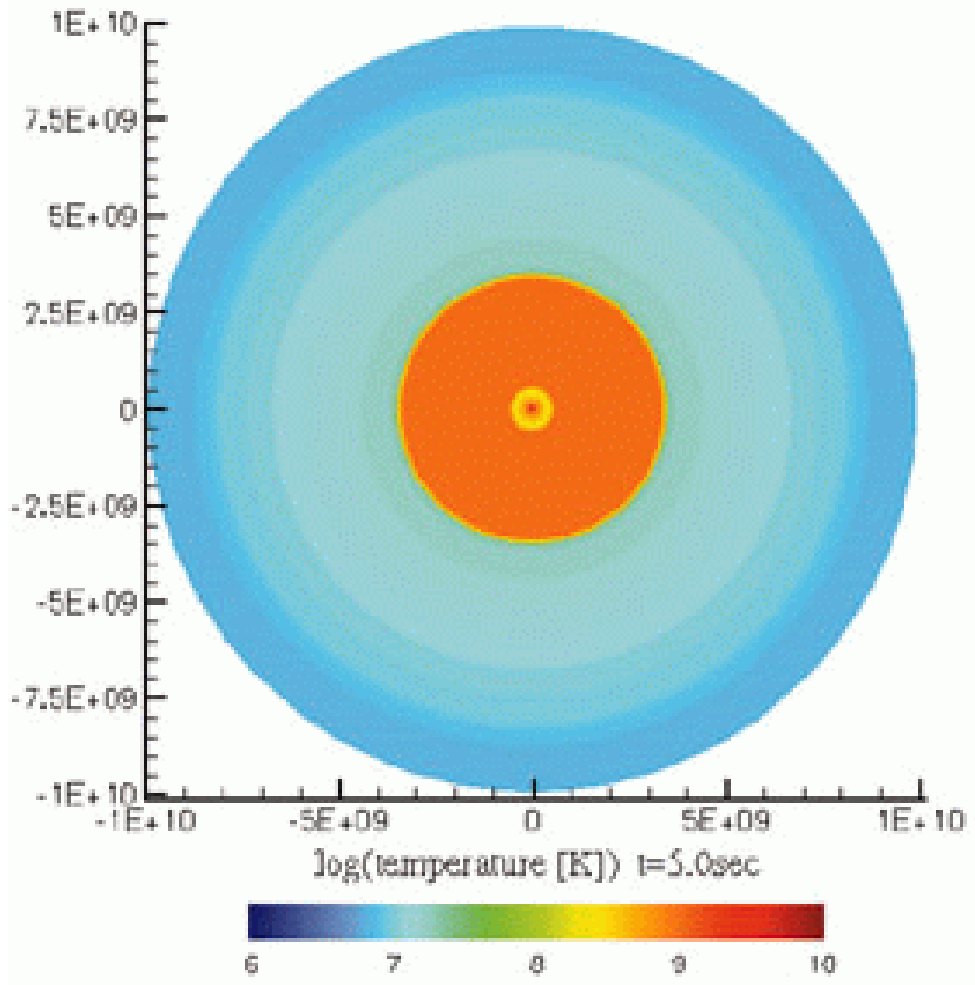}{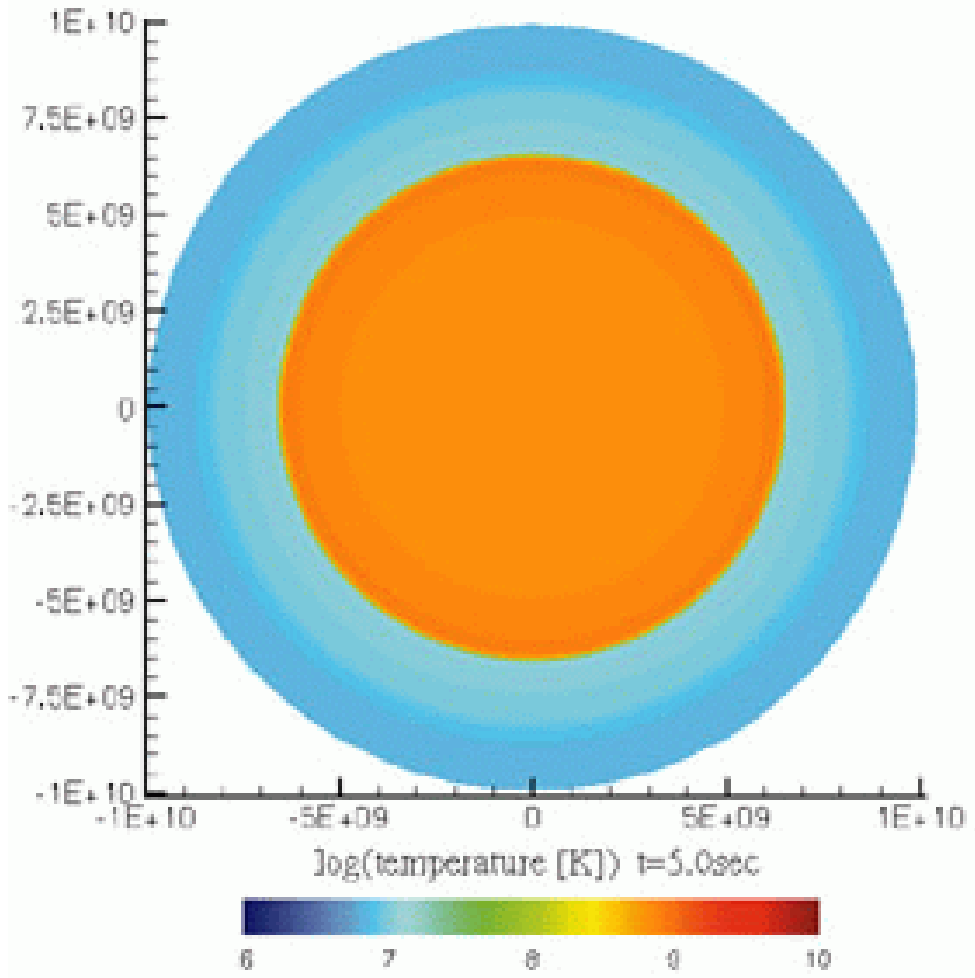}
\caption{Same as Fig.4, but for $t=5.0$ sec.
\label{fig6}}
\end{figure}

We first show the results of hydrodynamical calculations.
In Figs.~\ref{fig4},~\ref{fig5}, and~\ref{fig6}, we draw the contours of
temperature for the spherical explosion models at $t=1.0$ sec, 3.0 sec,
and 5.0 sec, respectively. Here we assumed the axial symmetry and equatorial
symmetry. The left panel corresponds to Model Sa and the right panel
to Model Sb. It is clearly found that the velocity of the shock
wave is faster in Model Sb than in Model Sa. This is because the energy density
behind the shock is greater in Model Sb, which means that
the pressure behind the shock is greater in Model Sb.

\begin{figure}
\plottwo{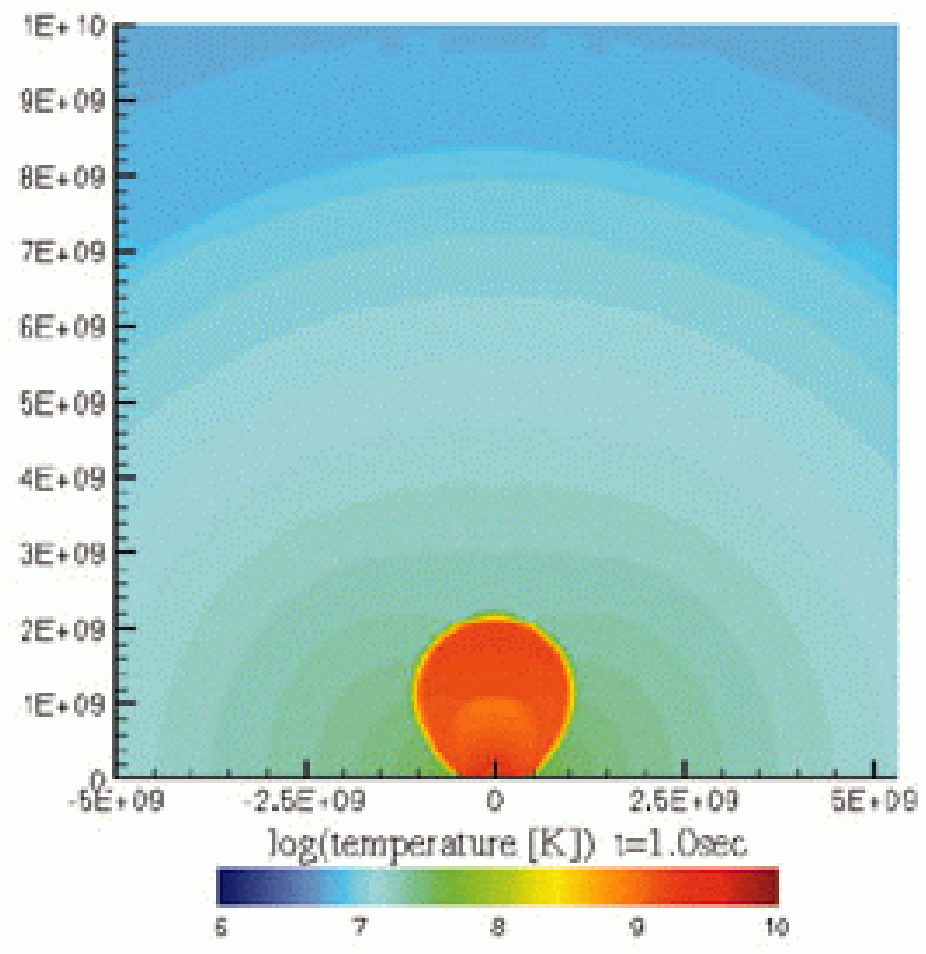}{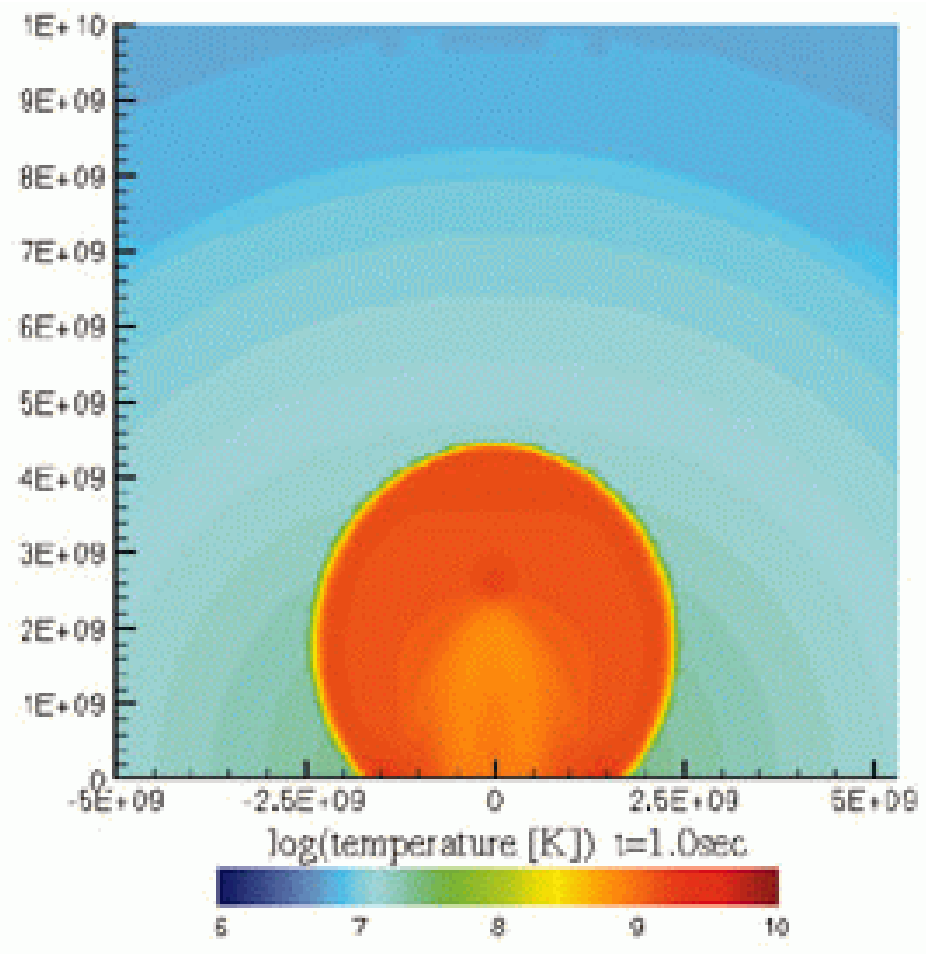}
\caption{Contours of temperature at $t=1.0$ sec. Left pannel corresponds
to Model Aa and right pannel corresponds to Model Ab. Only the northern
heimsphere is shown.
\label{fig7}}
\end{figure}

\begin{figure}
\plottwo{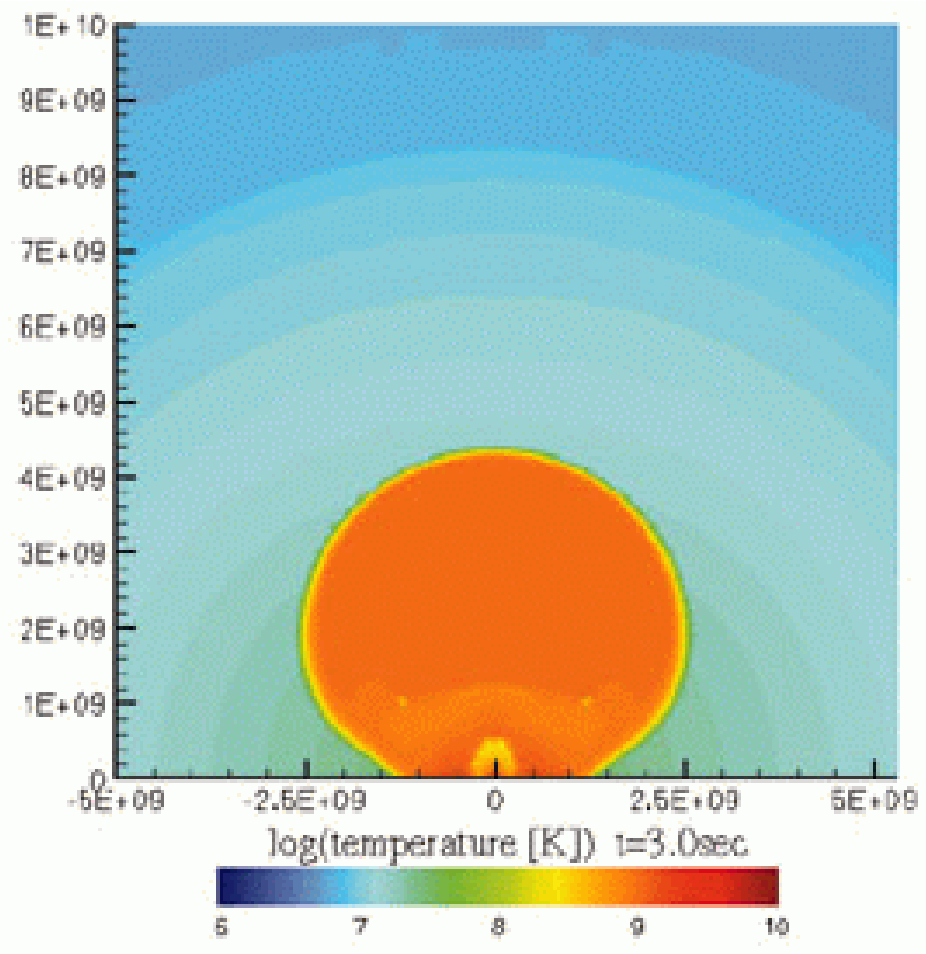}{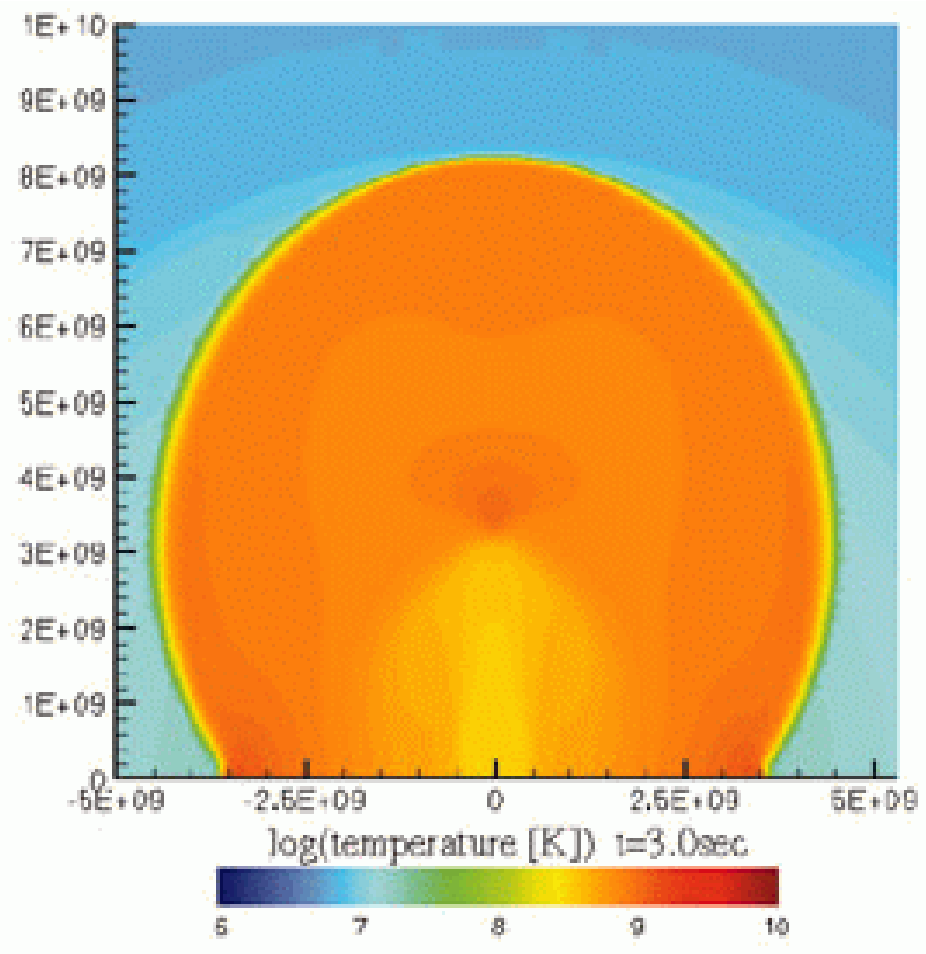}
\caption{Same as Fig.7, but for $t=3.0$ sec.
\label{fig8}}
\end{figure}

\begin{figure}
\plottwo{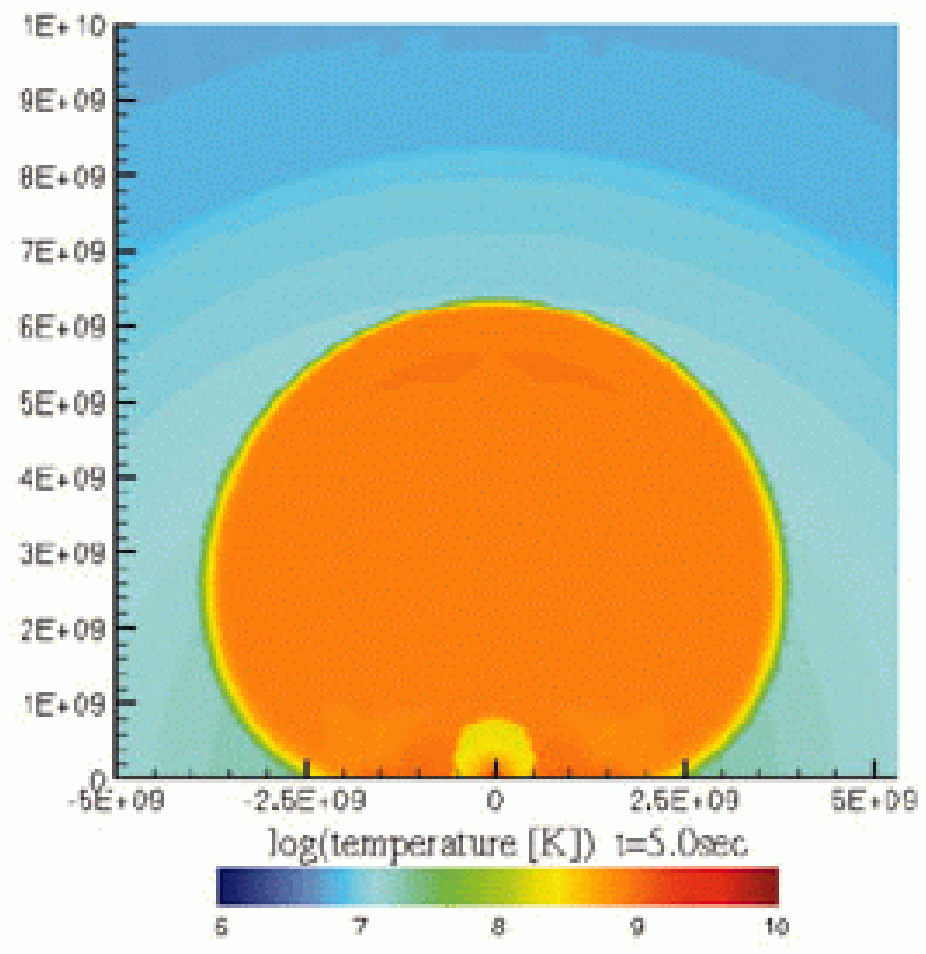}{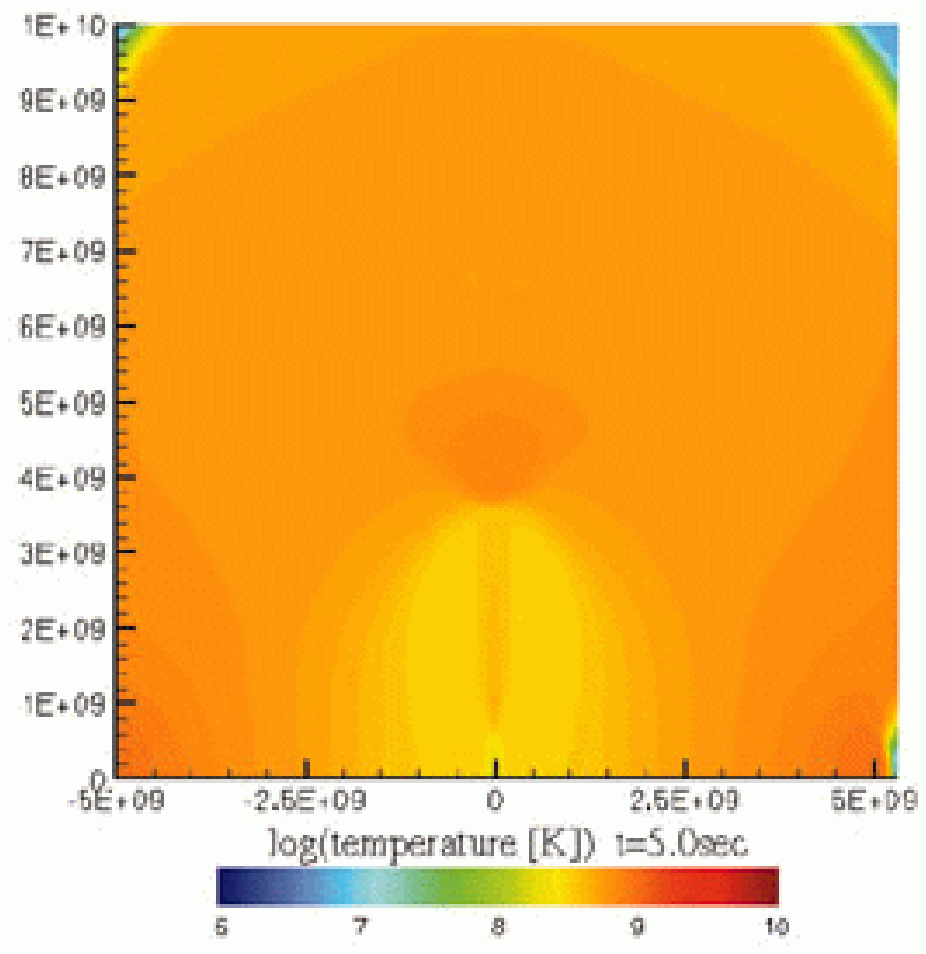}
\caption{Same as Fig.7, but for $t=5.0$ sec.
\label{fig9}}
\end{figure}

In Figs.~\ref{fig7},~\ref{fig8}, and~\ref{fig9}, we show the contours of
temperature for the asymmetric explosion models. In these figures, only the
upper hemisphere  is shown. The same tendency as in the spherical explosion
model, that the velocity of the shock wave is faster in Model Ab than in Aa, is
confirmed. It is also clearly shown that the velocity of the shock wave
depends on the zenith angle and jetlike explosion is induced, although
(as opposed to Maeda et al. 2002) only the thermal energy is injected. 
This reflects the initial density structure of the progenitor and is
consistent with MacFadyen and Woosley (1999).

\begin{figure}
\plottwo{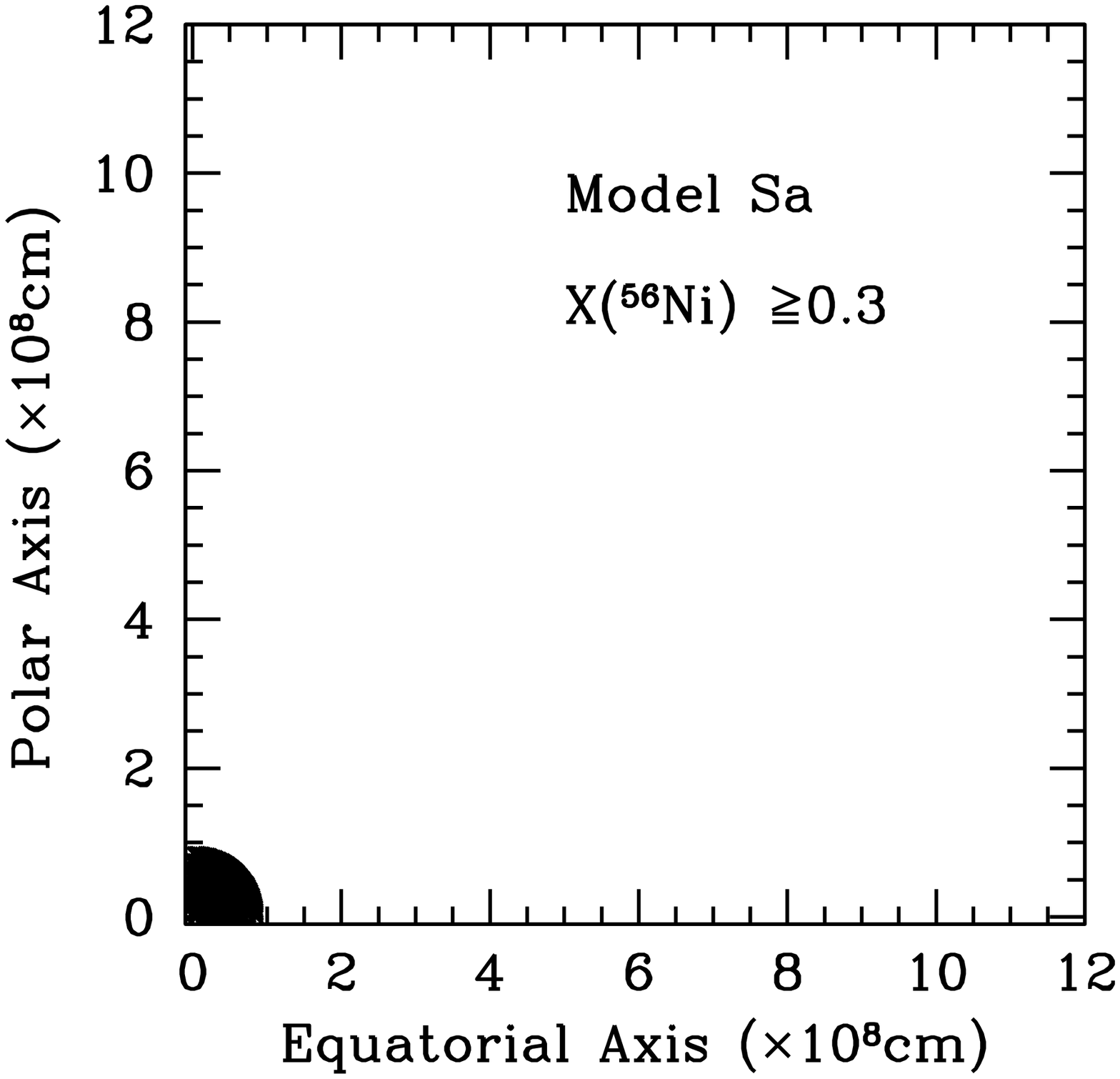}{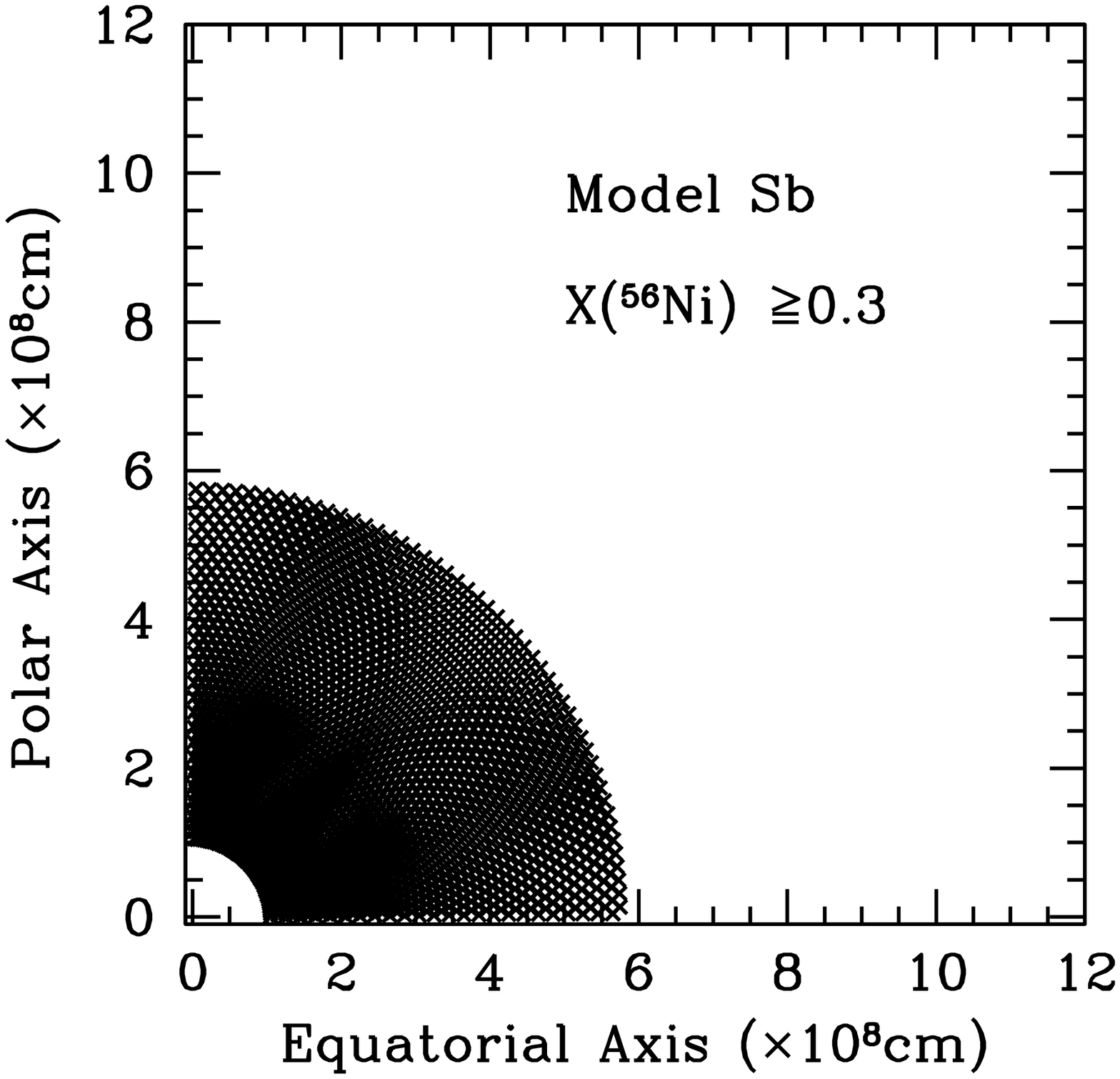}
\caption{Positions of the test particles at $t=0$ sec that meet the condition
that the mass fraction of $\rm ^{56}Ni$ becomes greater than 0.3.
Left pannel corresponds to Model Sa whereas right pannel corresponds
to Model Sb. 
\label{fig10}}
\end{figure}

\begin{figure}
\plottwo{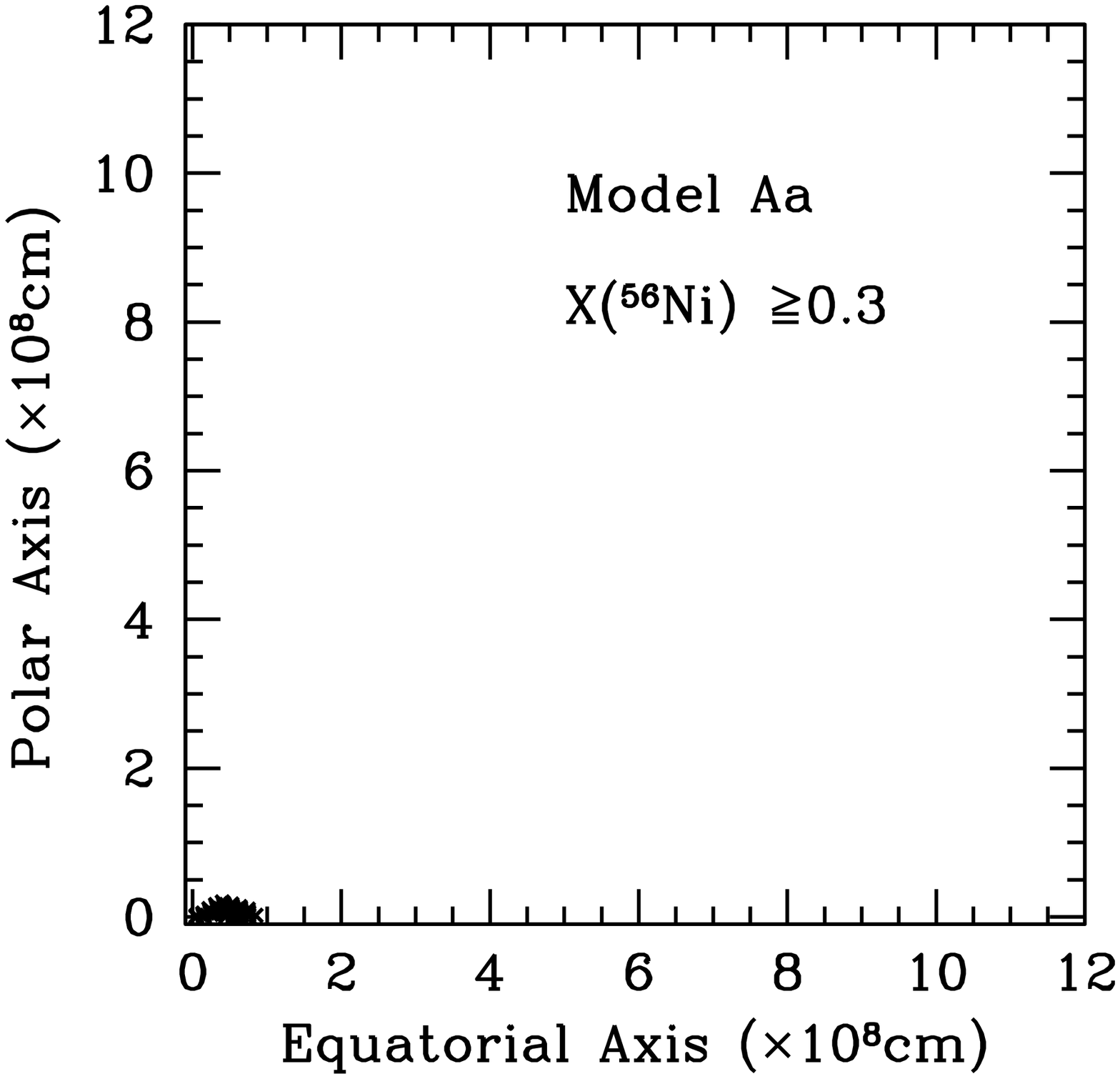}{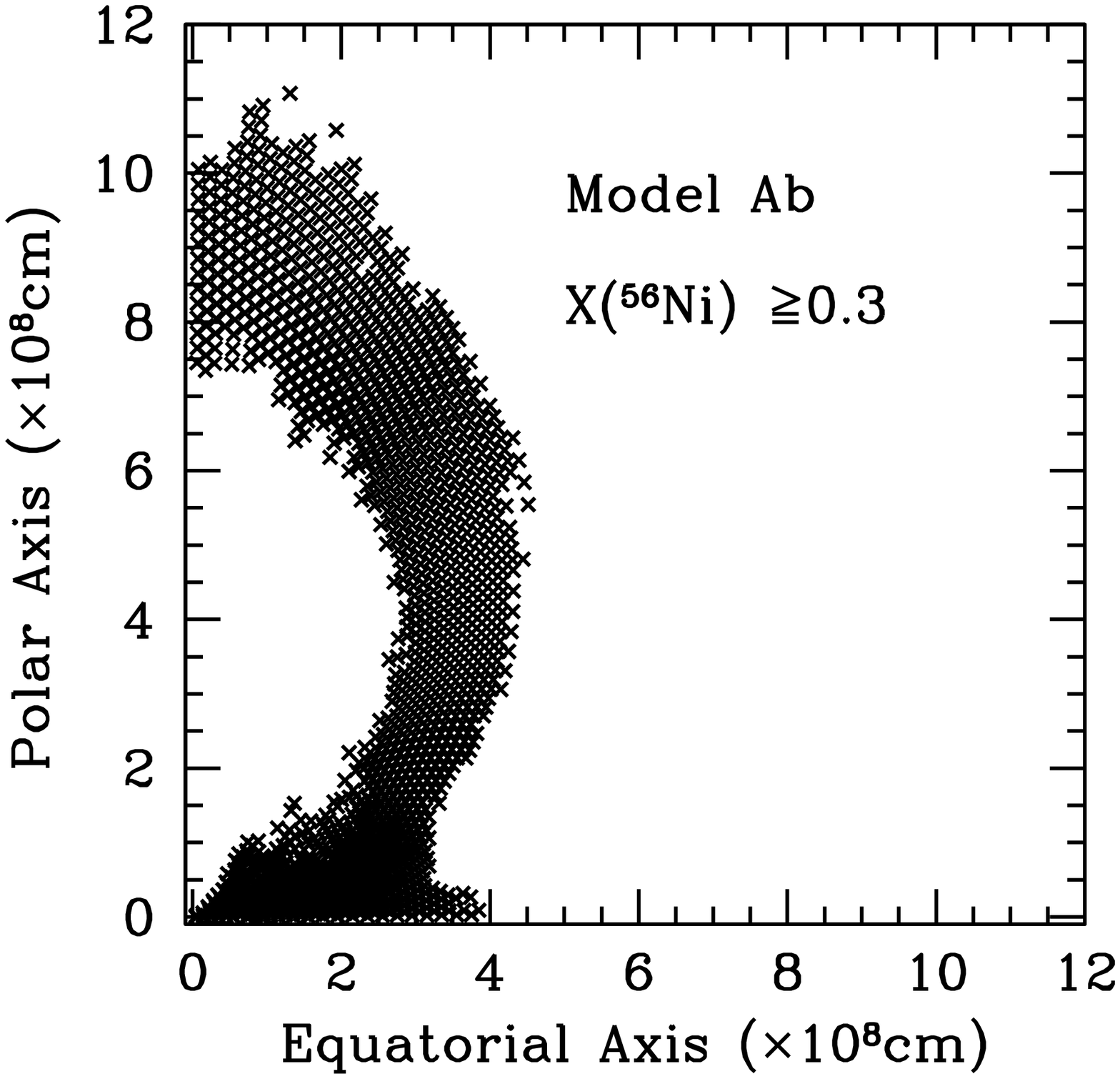}
\caption{Same with Fig.10 but for the asymmetric models.
Left pannel corresponds to Model Aa whereas right pannel corresponds
to Model Ab. 
\label{fig11}}
\end{figure}

\begin{figure}
\plotone{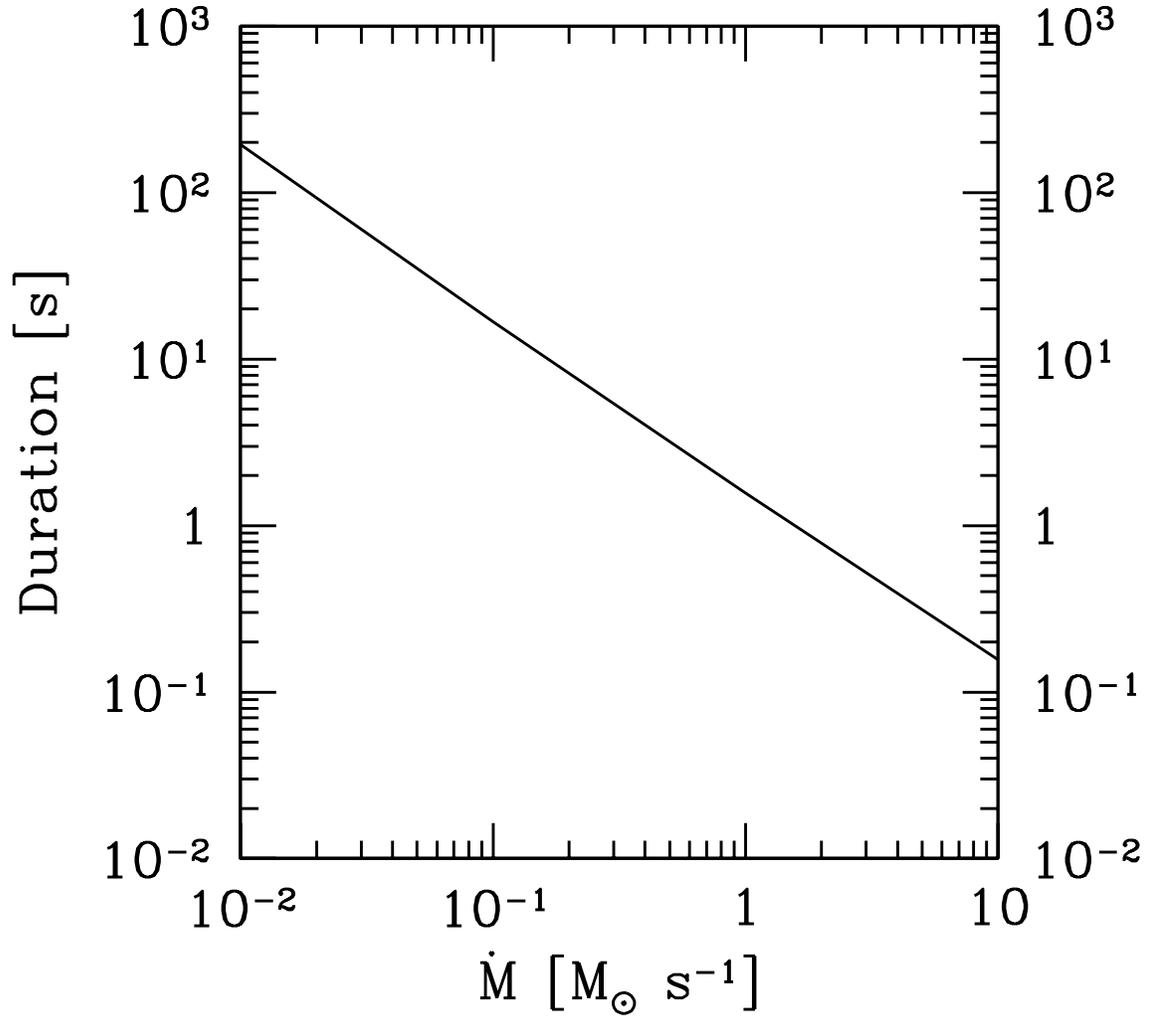}
\caption{
Duration of the neutrino emission from a collapsar as function
of the mass-accretion rate. The initial mass of the central black hole
is set to be 3$M_{\odot}$.
\label{fig12}}
\end{figure}

In Figs.~\ref{fig10} and~\ref{fig11}, we show the positions at $t=0$
of the test particles with the mass fraction of
$\rm ^{56}Ni$ greater than 0.3. It is clearly seen that there
is almost no region where $\rm ^{56}Ni$ is synthesized in Models Sa and Aa,
whereas $\rm ^{56}Ni$ is synthesized in large region for Models
Sb and Ab. In Table~\ref{tab3}, we summarize the abundance of heavy elements
in the ejecta for each model, assuming that all unstable nuclei produced in
the ejecta decay to the corresponding stable nuclei. The amount of
$\rm ^{56}Ni$ is also shown in the last line. From the table, we can
easily confirm that $\rm ^{56}Ni$, which decays to $\rm ^{56}Fe$, is hardly
synthesized in Models Sa and Aa (Note that almost all of $^{56}$Fe is generated
from the decay
of $\rm ^{56}Ni$). It is emphasized that the amount of $\rm ^{56}Ni$
synthesized in Models Sa and Aa is less than the required amount to explain
the luminosity of SN 1998bw ($\sim 0.7 M_{\odot}$; Iwamoto et al. 1998).
On the other hand, the amount of $\rm ^{56}Ni$ synthesized in Models Sb
and Ab is almost sufficient to explain the light curve of SN 1998bw.
It is noted that the progenitor before
the explosion is mainly composed of O, Ne, and Mg~\cite{nomoto88},
while the elements heavier than Si are synthesized through the explosive
nucleosynthesis~\cite{hashimoto95}. It is evident in Table~\ref{tab3}
that explosive nucleosynthesis occurs and the initial chemical composition
of the progenitor is changed to the heavier elements in Models Sb and Ab,
while nearly no explosive nucleosynthesis occurs in Models Sa and Aa.

Here we consider why iron elements are hardly synthesized
in Models Sa and Aa while much of them are synthesized in Models
Sb and Ab. The most significant parameter in explosive nucleosynthesis
in a massive star is the temperature. The criterion for the complete silicon
burning is $T_{\rm max} \ge 5\times 10^9$ [K]~\cite{thielemann96}. 
It is well known that the matter behind the shock wave is radiation
dominated and $T_{\rm max}$ can be well estimated by equating the
supernova (hypernova) energy with the radiation energy inside the radius $r$
of the shock front
\begin{eqnarray}
E_{\rm HN} = 10^{52} \left(\frac{E_{\rm HN}}{10^{52} \rm erg } 
\right) 
           = \frac{11 \pi^3}{45} \frac{k^4}{\hbar^3 c^3} r^3 T_{\rm max}^4
\;\;\; \rm [erg],
\end{eqnarray} 
where $E_{\rm HN}$ is the total explosion energy of hypernova,
$k$ is the Boltzmann constant, $\hbar$ is the Planck constant divided
by $2 \pi$, $c$ is the velocity of light. Here a spherical
explosion is assumed. This equation gives
\begin{eqnarray}
r = 5.7 \times 10^8 \left( \frac{5 \times 10^9 \rm K}{T_{\rm max}}   \right)^{4/3} \left( \frac{E_{\rm HN}}{10^{52} \rm erg} \right)^{1/3} \;\;\; \rm [cm].
\end{eqnarray} 
In the case of Model Sb, $\rm ^{56}Ni$ is synthesized within the edge for
the complete silicon burning ($\sim 5.7 \times 10^8$ cm).
In the case of Model Sa, on the other hand, matter start to move
outwards after the passage of
the shock wave, and almost all of the matter move away
($r \ge 6\times 10^8$ cm) before the injection of all the thermal energy
(= $10^{52}$ erg). This is the reason why almost no complete
silicon burning occurs in Model Sa. The situation should be the same in the
asymmetric explosion models. It is apparent from Table~\ref{tab3} that
the chemical composition of the ejecta depends more sensitively on the
timescale of thermal energy injection than on the asymmetry of the
initial density structure. It should be emphasized that the discussion
presented here can also be adopted to the pulsar-powered models for
GRBs~\cite{woosley02}, in which, as Woosley (2002) pointed out, very
little nickel will be produced unless a pulsar deposit at least $10^{51}$ erg
in a few tenths of a second. In the next section, we discuss the
implication of these results.

\section{Discussion}\label{discussion}

In this study, we assume that $\rm ^{56}Ni$ is synthesized
in the jet as done in Maeda et al. (2002). In this section, we
give further discussion under this assumption.

First, as stated in section~\ref{intro}, 
it is reported that no coincident SN is detected in GRB 010921
to a limit of 1.34 mag fainter than SN 1998bw at 99.7$\%$ confidence
is detected by Hubble
Space Telescope. If we simply interpret this result 
as a lack of $\rm ^{56}Ni$ in GRB 010921,
the amount of $\rm ^{56}Ni$ synthesized in GRB 010921
is less than 0.7$M_{\odot}$/(2.512$^{1.34}$)$\sim$0.2$M_{\odot}$,
which can be naturally explained by Models Sa and Aa. We want to emphasize
that the amount of $\rm ^{56}Ni$ in the ejecta can be small even if
the total explosion energy is much larger than that of
the normal collapse-driven
supernova, as long as the energy deposition rate is relatively low.
We insist that observations like GRB 010921 can be
naturally explained in the context of the type I collapsar model.

Second, we discuss the features of line emission in some X-ray afterglows.
As stated in section~\ref{intro}, strong iron K$\alpha$ emission
lines are found in X-ray afterglows~\citep{piro98,piro00,antonelli00,
yoshida01}.
There is also a report of emission lines of Mg, Si, S, Ar, and Ca with
an outflow velocity of order 0.1$c$ in the X-ray afterglow of
GRB 011211~\cite{reeves02}. According to Reeves et al. (2002),
the relative abundance of the light metals (Mg, Si, S, Ar, Ca) was found
to be $\sim 9$ times the solar value, whereas
the relative abundance of iron was found to be
$\le 1.4$ times the solar value.
Although the mechanism of such emission lines is still
controversial~\citep{paczynski98,lazzati99,rees00,vietri01,kotake01,kosenko02},
we want to point out that the chemical composition in the X-ray
afterglow may be naturally explained if the effects of the energy
deposition rate are taken into account, as shown in this study.
Here the origin of the emission lines is assumed to be the ejecta from
the underlying SN (HN). One may think that the relative abundance
of iron becomes small when the effect of fallback is taken into account.
However, if the jetlike explosion occurs to generate a GRB,
iron elements are synthesized around the jet axis as shown in Fig.~\ref{fig11}.
In this case, it will be hard to consider that the relative abundance of iron
becomes small due to the effects of the fallback.
According to Reeves et al. (2002), the velocity of the outflow which emits
the lines of heavy elements is of order 0.1$c$, which is about ten times
faster than the velocity of the ejecta in the normal collapse-driven 
supernova~\citep{nagataki98,nagataki00}. The required kinetic energy
is easily estimated as
\begin{eqnarray}
E_{\rm kin} = 9 \times 10^{51} \left(   \frac{M}{1M_{\odot}}     \right)
\left(    \frac{v}{0.1c}   \right)^2 \;\;\; \rm erg,
\end{eqnarray}
which means that the underlying SN is a hypernova.
We emphasize again that the abundance of the elements with intermediate
mass number such as Si and S becomes greater than that of iron even
if a hypernova explosion is adopted as long as the timescale of the energy
deposition is long.
If this discussion is true, there should be correlation between the chemical
composition of the ejecta and the timescale of the energy deposition.
This should mean that there should be correlation between the line
features in the X-ray afterglow and the duration of the GRB. 
If the emission lines are mainly composed of light elements such as Si and
S, the
duration of the GRB should be relatively long and vice versa.
In fact, it is reported that the duration of GRB 011211 is 270 seconds,
making it the longest burst ever observed by Beppo-SAX~\cite{reeves02}.
Although the interpretation of the line feature of X-ray afterglow of
GRB 011211 seems to be still controversial~\cite{rutledge02}, we hope
the increase of observations in near future.

Third, we consider that the observations of SN 1998bw can be well
explained by Models Sb and/or Ab, but not by Models Sa and Aa.
This means that
the energy deposition timescale should be relatively short ($\ll 10$ sec)
in order to explain the luminosity of SN 1998bw by the type I collapsar
model.
One may think that the amounts of $\rm ^{56}Ni$ produced
in Models Sb and Ab (0.14$M_{\odot}$ and 0.16$M_{\odot}$, respectively)
are still less than the observed value ($\sim
0.7M_{\odot}$; Iwamoto et al. 1998). However, the total explosion energy
is assumed to be $\sim 3\times 10^{52}$ ergs in Iwamoto et al. (1998)
while it is decreased to $10^{52}$ ergs in this study. Thus, we consider
that the observed amount can be explained with no problem when the
total explosion energy is set to be $\sim 3\times 10^{52}$ ergs in
Models Sb and/or Ab. If we believe the conclusion derived by Maeda et
al. (2002) that the line features in SN 1998bw can be well reproduced by
the jet-induced explosion model, Model Ab seems to be the best one.
However, we argue below that it seems to be
difficult even for the models Sb and Ab to generate a GRB. At present, we
are led to the conclusion that there seems to be
no model to generate a GRB
associated with a hypernova synthesizing much iron.

Now we discuss whether the results obtained in this study can 
constrain the model of the central engine of a hypernova accompanied by
a GRB. According to the scenario of the type I
collapsar model, the explosion energy larger than that of the normal
collapse-driven supernova may be given by the pair annihilation
of neutrinos from the accretion disk around the central black hole.
It takes about several seconds from the beginning of the core collapse
to form an accretion disk. In the meantime, the matter around the 
polar region falls directly into the central black hole. This is because the
matter around the polar region possess smaller angular momentum.
As a result, the polar region may become an adequate environment for
generating a fireball. The timescale to form a fireball is determined by the
timescale of neutrino emission from the accretion disk, which is
about 10 seconds~\cite{macfadyen99}. 
Although we pointed out in this study that the type I collapsar model can
naturally explain the observations of GRB 010921 and GRB 011211 in which
little amount of iron is required, this scenario may face
difficulty in producing a large amount of $\rm ^{56}Ni$
required in SN 1998bw,
since the energy deposition rate is too low.
If the energy deposition timescale is forced to be shorter ($\ll 10$ sec),
it seems to become doubtful whether a GRB can be really generated, as
explained below. To make the energy deposition
rate greater means the larger the mass accretion rate.
This
corresponds to the case in which the total angular momentum in the central
region of the progenitor is small. Then, it is doubted that
whether an asymmetric explosion occurs and an adequate environment
for generating a fireball is achieved around the polar region.
That is, we suspect that the density around the polar region can not become low
enough to realize a fireball when the mass accretion around the
equatorial plane is large. This is because there is little time for the 
mass around the polar region to fall into the black hole before the
explosion due to the pair annihilation of neutrinos occurs.

Let us investigate the above discussion quantitatively.
Popham et al. (1999) proposed an accretion-disk model including the
effect of general relativity as a collapsar
model~\cite{popham99}.
This model and the numerically computed collapsar
model~\cite{macfadyen99} are reproduced well by the analytical model of
the accretion disk by Nagataki et al. (2002).
The density, the
temperature and the disk thickness are fitted as follows
(see Nagataki et al. 2002 for details).
\begin{eqnarray}
    \label{eq:rho_N}
    \rho \left[ {\rm g \; cm^{-3}} \right]= 8.23 \times 10^{8}
    \left(\frac{M}{3M_{\odot}}\right)^{-1.7} 
    \left(\frac{\dot{M}}{0.1 M_{\odot} {\rm s}^{-1}}\right)^{1.03}
     \frac{1}{ \displaystyle{\left(\frac{r}{r_s}\right)^{1.07}
       \left( 1 + \left( \frac{r}{r_s} \right)  \right)^{0.76}}  } , \\ 
    \label{eq:temperature}
   T \left[ {\rm MeV} \right] = 2.3 \times \left(\frac{M}{3M_{\odot}}\right)^{-0.2}
   \left(\frac{\dot{M}}{0.1 M_{\odot} {\rm s}^{-1}}\right)^{0.108}
    \frac{1}{ \displaystyle{\left(\frac{r}{r_s}\right)^{0.425}
       \left( 1 + \left( \frac{r}{r_s} \right)  \right)^{0.21}}  } , \\  
    H \left[ {\rm cm} \right] = 5.8 \times 10^{6}  \left(\frac{M}{3M_{\odot}} \right)^{0.9}
\left(\frac{\dot{M}}{0.1 M_{\odot} {\rm s}^{-1}}\right)^{-0.0183}
 \frac{1}{\displaystyle{ \left(\frac{r}{r_s}\right)^{-1.66}
          \left( 1 + \left( \frac{r}{r_s} \right)  \right)^{0.3867}} },
\end{eqnarray}
where $M$ is the mass of the central black hole, $r$ is the radial
coordinate, and $r_s$ (= 10$^{7}$cm) is the core radius, respectively.
Note that the Schwarzschild radius is $\simeq 8.862{\rm km} (M/3
M_{\odot})$.
We see from Eq.~(\ref{eq:temperature}) that as the mass of the central
black hole grows, the temperature becomes lower, and the flux of
neutrinos decreases. In Fig.~\ref{fig12}, we show the duration
of neutrino emission. This (almost)
corresponds to the timescale in which the temperature at the
innermost region of the accretion disk becomes lower than $\sim$ 1
MeV. This will approximately reflect the timescale of GRB.
In this figure the initial mass of the central black hole
was set to be 3$M_{\odot}$. The time when
the temperature
of the inner-most region of the accretion disk becomes lower than 1 MeV
almost corresponds to the time when the mass of the black hole becomes $\sim
4M_{\odot}$.
From this figure, we find that the mass accretion rate has to be larger than
several $M_{\odot}$ s$^{-1}$ so that the energy deposition timescale
is shorter than 1 second. This mass accretion timescale is shorter
than that presented in MacFadyen and Woosley (1999). On the other hand,
if the mass accretion is so rapid, there will be no time to
make an adequate environment to produce a fireball around the polar
region as mentioned above.
Of course, we can not conclude that the type I collapsar model is
ruled out by the discussion of the explosive nucleosynthesis as a model
of hypernova associated with GRB.
We guess that there will be some ways to solve the difficulty presented
in this study. However, we believe that it is very important to point out
that the central engine of a GRB accompanied by SN (HN) explosion
can be constrained by the discussion of the explosive nucleosynthesis.
At present, we have to say that the Type I collapsar model is required
to be improved so as to explain the observed luminosity in SN 1998bw. 
It is noted that MacFadyen and Woosley (1999) pointed out the possibility
that a substantial amount of $\rm ^{56}Ni$ is produced in the accretion
disk and a part
of it is conveyed outwards by the viscosity-driven wind
(see their figure 16; see also~\cite{pruet03}). 
Their scenario will be important and it should be investigated whether
the behavior of the light curve and the line features of iron and oxygen
can be explained in such a framework
as well as the jetlike explosion model (Maeda et al. 2002).
In the study of Maeda et al. (2002), the narrow line of oxygen and
broad lines of iron can be explained when a jet-induced explosion
is assumed.

We consider that the variation of the energy deposition rate is naturally
caused by the variation of the progenitor's initial angular momentum.
Thus it may be still too early to consider that a GRB is always
accompanied by a hypernova that generates a large amount of iron elements. 
This is what we want to stress in this paper and we expect the increase
of observations in near future.

As mentioned in section~\ref{intro}, the accretion disk around the
central black hole may be the site of $\rm ^{56}Ni$ synthesis~\cite{pruet03}.
This scenario is widly supported to be most promising, although the
quantitative evaluation of the yield is remaining to be done.
If true, there may not be any correlation
between the duration of GRB and the amount of synthesized $\rm ^{56}Ni$
(the collapsar without the disk wind may make too little $\rm ^{56}Ni$ in
any case). Thus it might be possible to distinguish two scenarios
by future observations.

In this study, we used the nuclear reaction network which
contains 250 species (see table~\ref{tabnucl}). Thus we could not
discuss the nucleosynthesis of heavier elements such as in the r-process
nucleosynthesis. We consider that the polar region within the jet may be
an adequate site for r-process nucleosynthesis~\citep{macfadyen99,nagataki00,
nagataki01}. We are planning in near future to perform the calculation
of r-process nucleosynthesis in the context of jetlike explosion presented
in this study.

In this study, we modeled the jet launched by the neutrino heating
and a large opening angle of the jet was necessary for the jet
to produce a large enough amount of $\rm ^{56}Ni$ to explain the luminosity
of the
hypernova such as SN 1998bw. As stated in section~\ref{intro}, such a wide
jet will not make a GRB because a bulk Lorentz factor can not be
sufficiently large to
realize a fireball. Therefore, to realize a GRB in our scenario,
another narrow jet might be needed supposedly launched by MHD effect.
If the jet is launched by the MHD effect~\citep{blandford77}, a considerable
fraction of the explosion energy should be in the form of kinetic energy
and/or magnetic field energy~\cite{zhang03}. As a result the temperature should
be lower compared with the neutrino heating model until the reverse shock
converts the kinetic energy into the thermal energy~\cite{zhang03}.
Thus the explosive nucleosynthesis will be unlikely to occur in this
scenario unless the reverse shock propagates inwards immediately and makes
the temperature high enough to cause the explosive nucleosynthesis.

We made in this study an asymmetric initial progenitor model
by a simple prescriptism. This treatment should be, of course, improved.
We are planning to perform numerical calculations of core collapse in a
massive star that leads to a black hole formation including the effects
of gravity, rotation and neutrino transfer~\citep{yamada94,yamada97,kotake03}.
We will then investigate in detail the dependence of explosive
nucleosynthesis on the density structure at the time of the jet launch 
as a function of initial angular momentum in a progenitor
based on their results in the near future.

\section{Conclusion}\label{conclusion}

We have performed numerical calculations of explosive nucleosynthesis in
the context of the collapsar model using 2D relativistic 
hydrodynamical simulations. We have investigated the influence
of the structure of the progenitor and energy deposition rate on the
explosive nucleosynthesis, assuming that $\rm ^{56}Ni$ is mainly synthesized
in the jet. We have shown the amount of $\rm ^{56}Ni$
is very sensitive to the energy deposition rate rather than the structure
of the progenitor. We conclude that it is quite natural to detect
no underlying SN in some cases, such as GRB 010921. 
We have pointed out that the relative
abundance of the elements with intermediate mass number
such as Si and S in the X-ray afterglow
of GRB 011211 is naturally explained if the energy
deposition rate at the central engine is relatively small. 
We also predicted that there should be correlation between the line
features in the X-ray afterglow and the duration of the GRB.
Interestingly, the duration of GRB 011211 is 270 seconds,
the longest burst ever observed by Beppo-SAX although
it suffers from the effect of red-shift ($z_{\rm host}=2.14$). 
 Our results also imply that
the type I collapsar model in which the energy deposition rate is relatively
low ($\dot{E} \sim 10^{51}$ erg s$^{-1}$) as shown in MacFadyen and
Woosley (1999), may have difficulty in reproducing the observed amount of
$\rm ^{56}Ni$ in SN 1998bw. This means that the mechanism of the
central engine of hypernova and GRB is constrained by the discussion
of explosive nucleosynthesis. 
We hope that the increase of observations will lead to the improvement
of theories on GRB in near future.

\acknowledgments
The authors thank to Dr. Blinnikov for useful comments.
S.N. thanks the Yukawa Institute for Theoretical Physics at
Kyoto University, where this work was initiated during the YITP-W-01-06 on
'GRB2001' and completed during the YITP-W-99-99
on 'Blackholes, Gravitational Lens, and Gamma-Ray Bursts'.
Hydrodynamic calculations were carried out on
NEC SX-5 system at Cybermedia Center, Osaka University. 
This research has been supported in part by the Grant-in-Aid by the
Ministry of Education, Culture, Sports, Science and Technology of
Japan (No.14079202, No.S14102004, No.S14740166).



\clearpage

\clearpage

\begin{table*}
\begin{center}
\begin{tabular}{ccccccccccccccccc}
\tableline
\tableline
      &                 &  $\dot{E}$      &  $E_{\rm tot}$ \\
Model &  progenitor     &  [erg s$^{-1}$] &  [erg]          \\
\tableline
Sa    & spherical       &    $10^{51}$    & $10^{52}$       \\
Sb    & spherical       &    $\infty$     & $10^{52}$       \\
Aa    & asymmetric      &    $10^{51}$    & $10^{52}$       \\
Ab    & asymmetric      &    $\infty$     & $10^{52}$       \\
\tableline
\end{tabular}
\tablenum{1}
\caption{
Models, structure of the progenitor, thermal energy deposition rate
(erg s$^{-1}$), and total explosion energy. See detail in section 2.1.2.
}\label{tab1}
\end{center}
\end{table*}

\begin{table*}
\begin{center}
\begin{tabular}{ccccccccccccccccc}
\tableline
\tableline
Element & $\rm A_{min}$ & $\rm A_{max}$  & Element & $\rm A_{min}$ &
$\rm A_{max}$ & Element & $\rm A_{min}$ & $\rm A_{max}$ \\
\tableline
N & 1 & 1 & Al& 24 & 30 & V & 44 & 54 \\
H & 1 & 1 & Si& 26 & 33 & Cr& 46 & 55 \\
He& 4 & 4 & P & 28 & 36 & Mn& 48 & 58 \\
C & 11& 14& S & 31 & 37 & Fe& 52 & 61 \\
N & 12& 15& Cl& 32 & 40 & Co& 54 & 64 \\
O & 14& 19& Ar& 35 & 45 & Ni& 56 & 65 \\
F & 17& 22& K & 36 & 48 & Cu& 58 & 68 \\
Ne& 18& 23& Ca& 39 & 49 & Zn& 60 & 71 \\
Na& 20& 26& Sc& 40 & 51 & Ga& 62 & 73 \\
Mg& 22& 27& Ti& 42 & 52 & Ge& 64 & 74 \\
\tableline
\end{tabular}
\tablenum{2}
\caption{
Nuclear Reaction Network Employed
}\label{tabnucl}
\end{center}
\end{table*}

\begin{table*}
\begin{center}
\begin{tabular}{ccccccccccccccccc}
\tableline
\tableline
Element & Model Sa & Model Sb  & Model Aa & Model Ab \\
\tableline
O  & 9.3E+0 & 8.8E+0 & 9.1E+0 & 8.5E+0 \\
Mg & 5.4E-1 & 5.0E-1 & 5.3E-1 & 4.9E-1 \\
Si & 2.4E-1 & 3.8E-1 & 2.4E-1 & 3.1E-1 \\
S  & 3.3E-2 & 1.2E-1 & 3.4E-2 & 9.0E-2 \\
Ar & 6.6E-3 & 2.4E-2 & 7.7E-3 & 1.8E-2 \\
Ca & 4.5E-3 & 1.9E-2 & 5.6E-3 & 1.7E-2 \\
Ti & 1.2E-5 & 9.4E-4 & 8.3E-6 & 1.0E-3 \\
Fe & 1.4E-2 & 3.1E-1 & 8.0E-2 & 3.7E-1 \\
$^{56}$Ni   & 6.1E-3 & 1.4E-1 & 2.7E-2 & 1.6E-1\\
\tableline
\end{tabular}
\tablenum{3}
\caption{
Abundance of heavy elements in the ejecta for each model in units of
$M_{\odot}$. All unstable nuclei produced in the ejecta are assumed to
decay to the corresponding stable nuclei. The amount of $\rm ^{56}Ni$
is also shown in the last row.
}\label{tab3}
\end{center}
\end{table*}

\end{document}